\documentclass[final,5p,times,twocolumn]{elsarticle}

\usepackage{graphicx}
\usepackage{amssymb}
\usepackage{amsmath}
\usepackage{empheq}
\usepackage{comment}
\usepackage{subcaption}
\usepackage{balance}
\usepackage{placeins}
\usepackage{enumerate}
\usepackage[inline,shortlabels]{enumitem}
\usepackage[usenames,dvipsnames]{color}
\usepackage[colorlinks=true,linkcolor=Maroon,citecolor=Maroon,urlcolor=Maroon]{hyperref} 
\usepackage{url}

\usepackage[usenames,dvipsnames]{color} 
\usepackage{floatpag} 					
\usepackage{footmisc} 					
\usepackage{booktabs}                   
\usepackage{balance}                    
\usepackage{soul}                       

\makeatletter
\def\ps@pprintTitle{%
 \let\@oddhead\@empty 
 \let\@evenhead\@empty
 \def\@oddfoot{}%
 \let\@evenfoot\@oddfoot}
 \makeatother

\begin{document}
\begin{frontmatter}

\title{Early Career Citations Capture Judicial Idiosyncrasies and Predict Judgments}

\author[add1,add2]{$^\dagger$Robert Mahari}
\author[add1,add3]{$^\dagger$Sandro Claudio Lera$^*$}

\address[add1]{\scriptsize Massachusetts Institute of Technology, Cambridge, USA}
\address[add2]{\scriptsize Harvard Law School, Cambridge, USA}
\address[add3]{\scriptsize Southern University of Science and Technology, Shenzhen, China}

\begin{abstract}

Judicial impartiality is a cornerstone of well-functioning legal systems. We assemble a dataset of 112,312 civil lawsuits in U.S. District Courts to study the effect of extraneous factors on judicial decision making. We show that cases are randomly assigned to judges and that biographical judge features are predictive of judicial decisions. We use low-dimensional representations of judges' early-career citation records as generic representations of judicial idiosyncrasies. These predict future judgments with accuracies exceeding 65\% for high-confidence predictions on balanced out-of-sample test cases. For 6-8\% of judges, these representations are significant predictors across all judgments. These findings indicate that a small but significant group of judges routinely relies on extraneous factors and careful vetting of judges prior to appointment may partially address this issue. Our use of low-dimensional representations of citation records may also be generalized to other jurisdictions or to study other aspects of judicial decision making.

\vspace{1cm}

\end{abstract}
\end{frontmatter}

\renewcommand*{\thefootnote}{\fnsymbol{footnote}} 
\footnotetext[0]{$^\dagger$Both authors contributed equally.}
\footnotetext[0]{$^*$Corresponding author: \texttt{slera@mit.edu}}


Judicial impartiality is the principle that justice should be based on objective criteria and that judges should not be swayed by their own idiosyncrasies.
Impartiality has been a pillar of judicial systems for thousands of years: 
it is featured in the Magna Carta~\cite{mckechnie2022magna}, 
enshrined in religious texts~\cite{schiffman2008chapter}, 
memorialized in the Universal Declaration of Human Rights,
and represented by the blindfold worn by Lady Justice~\cite{curtis1986images}.
Though they are expected to be even-handed, judges are human and thus not immune to partiality. 
As a result, judicial systems have been designed to minimize the effect of extraneous influences on judges.
For example, in the U.S. Federal Judiciary, judges serve life-time appointments, are subject to ethics rules, and assigned cases at random. 

Our analysis of judicial impartiality builds on a voluminous literature on the scope of permissible judicial discretion.
It is inevitable that judges will be faced with situations that are not clearly spelled out in the law and must apply some extraneous knowledge, intuition, or philosophy to come to a resolution.
Legal philosophers have disagreed greatly on what judges should do when faced with ambiguity, including: 
prioritizing legally fit interpretations based on their accord with the political ideals a judge thinks are correct~\cite{dworkin1981natural}, 
adhering strictly to legal texts and precedents~\cite{nelson2005textualism}, 
seeking to take into account the position of the executive branch~\cite{sunstein2005beyond}, 
emphasizing legal rules and the system's internal logic~\cite{hart1961concept}, 
or insisting on the law's morality and coherence~\cite{fuller1964morality}.
Some of scholars segment lawsuits into ``easy'' cases, where a judgement may be deduced by applying law to facts in a ``legalistic'' manner, and ``hard cases'', where discretion might come into play~\cite{marmor1990no}.
Many jurists, with different perspectives on judicial philosophy, agree that the vast majority of cases, at least 90\%, are easy~\cite{howard2014courts, posner2010some, cardozo1924growth}.
With this background in mind, we believe that our large-scale quantitative study can help shed empirical light on several questions related to judicial impartiality and discretion.
First, how frequently are judges' decisions based on extraneous factors? 
If judges only apply their own philosophies to resolve relatively rare ambiguous situations, then one would expect this to happen infrequently, especially at lower court levels where most disputes should be of a routine nature.
Second, are there some judges who appear to frequently rely on extraneous factors in making their decisions?
While some cases may necessarily require a judge to exercise discretion, the random assignment of cases to judges should make it unlikely for a judge to be regularly faced with these legally ambiguous cases.
Thus, as long as cases are assigned randomly, judges who are systematically making decisions based on non-case related information may be displaying partiality.

Given their societal importance, the influence of judicial idiosyncrasies has been broadly studied.
Prior work has found that judicial decisions are affected by extraneous factors such as 
the time of day that a case is brought~\cite{danziger2011extraneous}, 
racial biases~\cite{arnold2018racial}, 
and gender biases~\cite{eisenberg2012does}.
By contrast, judges in Indian criminal courts were shown to display little in-group bias related to religion and ethnicity~\cite{bhowmick2021group}. 

One approach to studying judicial impartiality is to predict decisions on the basis of extraneous factors.
The successful prediction of judgments based on variables that should be irrelevant can serve to highlight discretionary decisions. 
Beyond its academic appeal, such predictions are practically valuable since litigants, law firms, activists, insurance companies, and litigation funders all have strong interests in predicting judgments~\cite{Lera2022}.
Computational judgment prediction has also been proposed as a way to help identify ``correct'' legal outcomes on the basis of case facts and laws.
Research on the computational prediction of judicial decision dates back to at least the 1960s~\cite{lawlor1962information}.
Since then, several studies have sought to predict decisions in the U.S. Supreme Court, where data is particularly accessible and decisions are highly consequential~\cite{ruger2004supreme}.
Models of U.S. Supreme Court judgments blending chronological variables, case background variables, lower court decisions, and extraneous justice-specific variables have been shown to successfully predict decisions~\cite{sharma2015using,katz2017general}.
Meanwhile, judgment prediction has emerged as a natural language processing task that seeks to predict judgments based on text about case facts and laws~\cite{zhong2018legal, medvedeva2020using, chalkidis2019neural}. 
The use of prediction to identify correct legal outcomes has also been proposed as a way to mitigate judicial partiality~\cite{kleinberg2018human}. 
Yet, some jurisdictions---among them France---have outlawed the use of algorithmic tools to predict or evaluate judicial behavior~\cite{g2022ai}.

One of the primary obstacles for the study of judicial impartiality is data availability~\cite{pah2020build}.
As a result, most existing studies either focus on higher courts where data is available, like the U.S. Supreme Court, or on a small number of lower courts for which data could be obtained.
Some prior studies have also been criticized for their small datasets or statistical methodologies~\cite{glockner2016irrational}.
Other studies acknowledge the challenges for using random case assignment to study the effect of judges such as data access and effects that might undermine random assignment, especially when one focuses only on a specific court~\cite{eisenberg2012does}.
While U.S. Supreme Court decisions normally relate to consequential societal issues, run-of-the-mill legal problems tend to be decided by lower courts.
In the U.S. Federal System, District Courts handle many more cases than higher courts and so they play a critical role in interpreting and applying laws.\footnote{
To appear in Federal District Court, a claim must exceed \$75,000 \emph{and} the parties must be citizens of different states (28 U.S.C. § 1332) or the legal dispute must involve a question of federal law (28 U.S.C. § 1331).
}
Despite their importance, systematic quantitative research on judge's reliance on extraneous factors across District Courts has been lacking due to limited access to data. 

In this study, we overcome the data availability challenge by merging several datasets and applying natural language processing (NLP) techniques to obtain information on $112,312$ civil U.S. District Court cases dating back to 1880.
We use this new dataset to conduct a large-scale systematic analysis of judicial impartiality.
Leveraging the observed case type, we confirm that cases are randomly assigned to judges.
We demonstrate that, despite this random case assignment, judges' biographical characteristics---such as gender, party affiliation, and past decisions---are predictive of their judgments.
We achieve prediction accuracies of up to 56\% - 65\% on balanced out-of-sample test cases, depending on the case type. 
In particular, we find that judges' historical win rates are highly predictive of their future decisions.
Since the U.S. is a common law system, judges base their decisions on citations to past cases.
These citation decisions thus provide insights into how judges reason. 
We show that the observed idiosyncrasies can be captured through judges' citation records and that these tendencies develop early in judges' careers.
Concretely, we demonstrate that low dimensional representations of judges' early-career citation records may be used to predict the outcome of future cases. 
Finally, we show that for around 6-8\% of U.S. District Court judges, judgments can be systematically predicted with statistical significance solely based on their early-career citation decisions.
Our study thus provides large-scale evidence that U.S. District Court judges are affected by extraneous factors and offers a generalizable set of methods to study different types of judicial idiosyncrasies in any common law jurisdiction.
While this work does not prove that any specific judge is biased, as this would require a fuller context-specific inquiry, it suggests that a small but significant group of judges systematically relies on extraneous factors.
These results are relevant to legal reform and practice. 
Our findings offer quantitative evidence to guide judicial reform and to recognize and mitigate potential partiality among judges. 
This information can also identify cases with particularly high or low winning probabilities, with important implications for litigators and prospective litigants as well as industries like insurance and litigation finance.

\begin{figure}[!htb]
    \centering
    \includegraphics[width=0.5\textwidth]{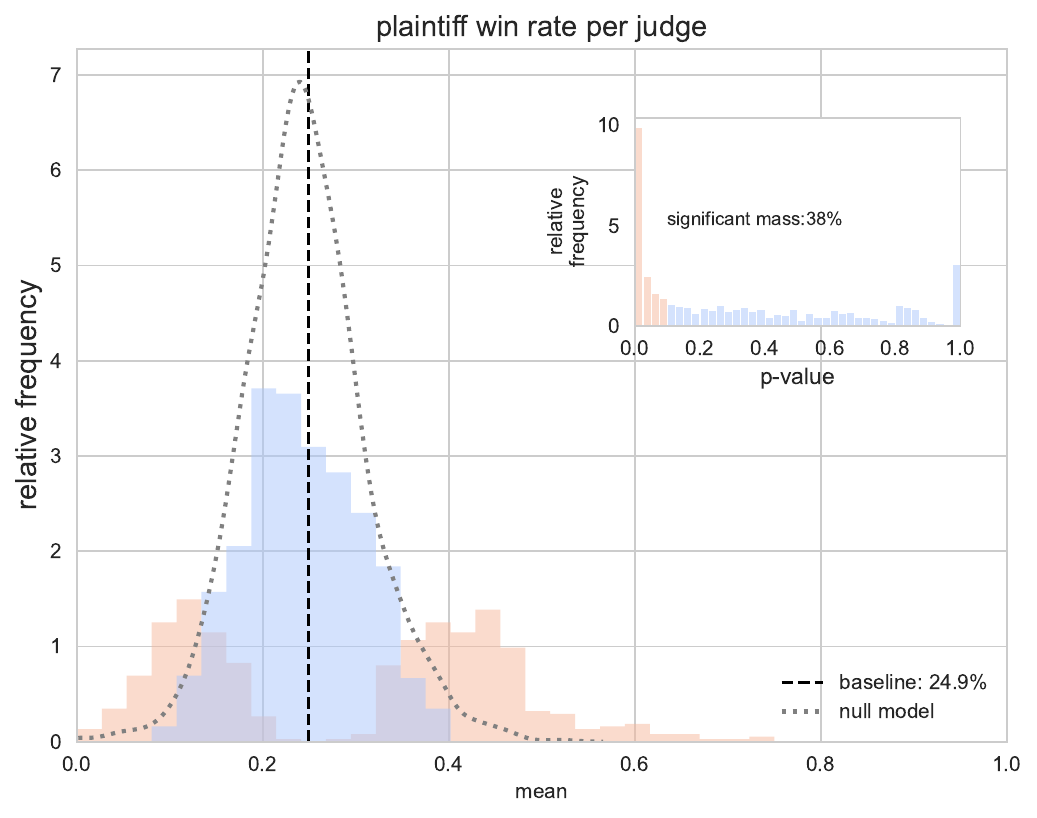}
    \caption{
            \textbf{Distribution of plaintiff win rates across judges}.
            Red bars correspond to win rates that deviate significantly from the baseline.
            Significance is measured at the 10\% level of a two-sided binomial test.
            The dotted line shows the distribution of win rates sampled from the binomial null-model $B(p=0.25,n)$.             
            (inset)
            Associated distribution of p-values. 
            38\% of all judges deviate from the baseline in a statistically significant manner. 
            This is more than three times higher than the rate of false positives one would expect at the 10\% level. 
            }
    \label{fig:distribution_of_win_rates}
\end{figure}

\section*{Results}

Prior studies on judicial impartiality have been limited by data availability.
We overcome this challenge by assembling a novel dataset of U.S. District Court decisions from 1880 onwards.
We integrate multiple existing datasets and augment them using NLP techniques (see Methods).
This dataset enables, to the best of our knowledge, the first large-scale systematic analysis of judicial impartiality across all 94 courts in the U.S. District Court System.

\subsection*{Confirming Random Case Assignment}
\label{sec:confirming_randomness}

The random assignment of cases to judges has long been viewed as a prerequisite for impartial justice.
From a quantitative perspective, random assignment is an important assumption because without it, any type of observed reliance on extraneous factors could plausibly be attributed to judges' preferences for certain cases.
Random case assignment has thus featured prominently in many previous studies on courts and judicial behavior~\cite{kleinberg2018human, kling2006incarceration, di2013criminal, harding2017short, collinson2023eviction}.
Here we conduct a large-scale investigation of random assignment and find that cases are indeed randomly assigned in U.S. District Courts.

U.S. District Courts explicitly state that cases must be assigned to judges at random.
The strategies employed to maintain this randomness vary from court to court.
A pertinent example is the Case Assignment System employed by the U.S District Court for the District of Minnesota:
This system uses a virtual deck of cards, where each card bears the name of an available judge. 
To assign a judge to a new case, a card is randomly drawn from the deck.
However, not all U.S. District Courts publish the detailed procedure they follow to ensure random assignment.
Meanwhile, most existing literature that relies on random assignment of judges tends to concentrate on specific courts rather than a court system as a whole. 
In particular, prior analyses of case assignment have usually been confined to a small number of courts~\cite{leslie2017unintended, dobbie2018effects}.
We conduct a large-scale test of random assignment across all U.S. District Courts based on the observed case type going back to the 1880s.
Concretely, our analysis encompasses the evaluation of 18,355 datasets.
Each dataset represents a set of case assignments of a given case type to a given judge during a certain decade. 
For each such dataset, we test the null hypothesis that posits random assignment of cases.
Only a negligible 0.7\% of these datasets demonstrate deviations from the expected random distribution (see Methods).

To investigate potential subtle biases in case assignments, we further categorize cases into four groups based on the litigants' identities: governmental entities, corporations, private individuals, or a mix of party types. 
After grouping cases by jurisdiction, decade, and judge, only 0.4\% of hypotheses are rejected (see SI Appendix).

These results thus substantiate the assertion that case assignment is random and lends credibility to the underlying premise of studies such as ours. 

\subsection*{Observing Reliance on Extraneous Factors in Judicial Decisions}
\label{sec:observed_bias}

Conditional on random case assignment, 
we expect plaintiff win rates averaged over a given impartial judge's career to be similar to the average plaintiff win rate across all judges (approximately $25\%$ across all cases in our data).
More formally, we expect that, for any given impartial judge, the plaintiff win rate will follow a binomial distribution, $B(p,n)$, with $p=0.25$ and $n$ equal to the number of cases that have been adjudicated by the judge. 
Our study is based on a novel dataset of 2,394 U.S. Federal  District Court judges and their decisions in 112,312 cases. 
In contrast to the binomial baseline, we observe that almost $40\%$ of all judges deviate in a statistically significant manner from the baseline (Figure \ref{fig:distribution_of_win_rates}).

Two non-competing explanations emerge for this observation: 
either confounding variables such as the geographic location (circuit) or time (case filing date) impact win rates of otherwise impartial judges, 
or some judges systematically favor or disfavor plaintiffs. 
Our study provides evidence for both of these explanations.

\subsection*{Prediction with Biographic Information}
\label{sec:bio_prediction}

Prompted by the large variance in the plaintiff win rate distribution, we examine whether biographic information about judges' personal and professional backgrounds can be used to predict their judgments.
To this end, we train binary classifiers that predict judgments based on biographic judge features.
Building on prior work on judicial decision making~\cite{kleinberg2018human}, we train gradient boost classifiers that can capture non-linear relationships.
However, qualitatively similar results are obtained with a logistic regression with Ridge penalty, with a multilayer perceptron and a random forest (see SI Appendix).
For each case, two information categories are available:
First, case-specific confounding variables, which include the case decision date, case type, and judicial circuit. 
Second, judge-specific information, which encompasses the judge's party affiliation and gender as fixed effects as well as the average win rate for the judge's previously adjudicated cases, promotional status, experience, and workload (see SI Appendix for summary statistics).
A judge's party affiliation is inferred based on the party of the U.S. President that appointed that judge.
The promotional status captures whether a judge has been promoted from a  District Court to another court. 
While the judicial system is generally designed to be free of external incentives,
the ability to be promoted to a higher court is an obvious external incentive that may affect how judges behave~\cite{epstein2012behavior}.
The workload and experience are calculated up to the relevant decision date (see Methods for details).
In principle, none of these judge-specific features are legally relevant and none of them should have a bearing on case outcomes.

\begin{figure}[!htb]
    \centering
    \includegraphics[width=0.5\textwidth]{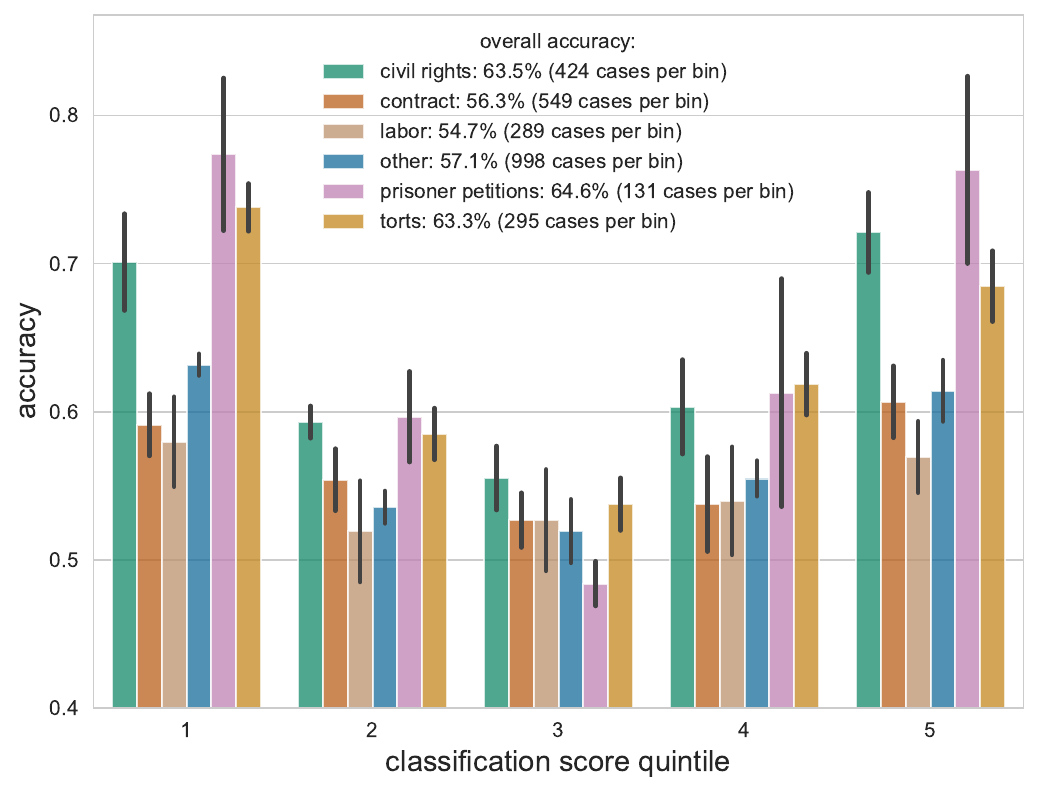}
    \caption{\textbf{Accuracy of the gradient boost probabilistic classifier that predicts case outcome.} The model was trained with biographic judge features and we report the accuracy per confidence score quantile for each case type.}
    \label{fig:biographic_feature_classification_results}
\end{figure}

\begin{figure}[!htb]
    \centering
    \includegraphics[width=0.5\textwidth]{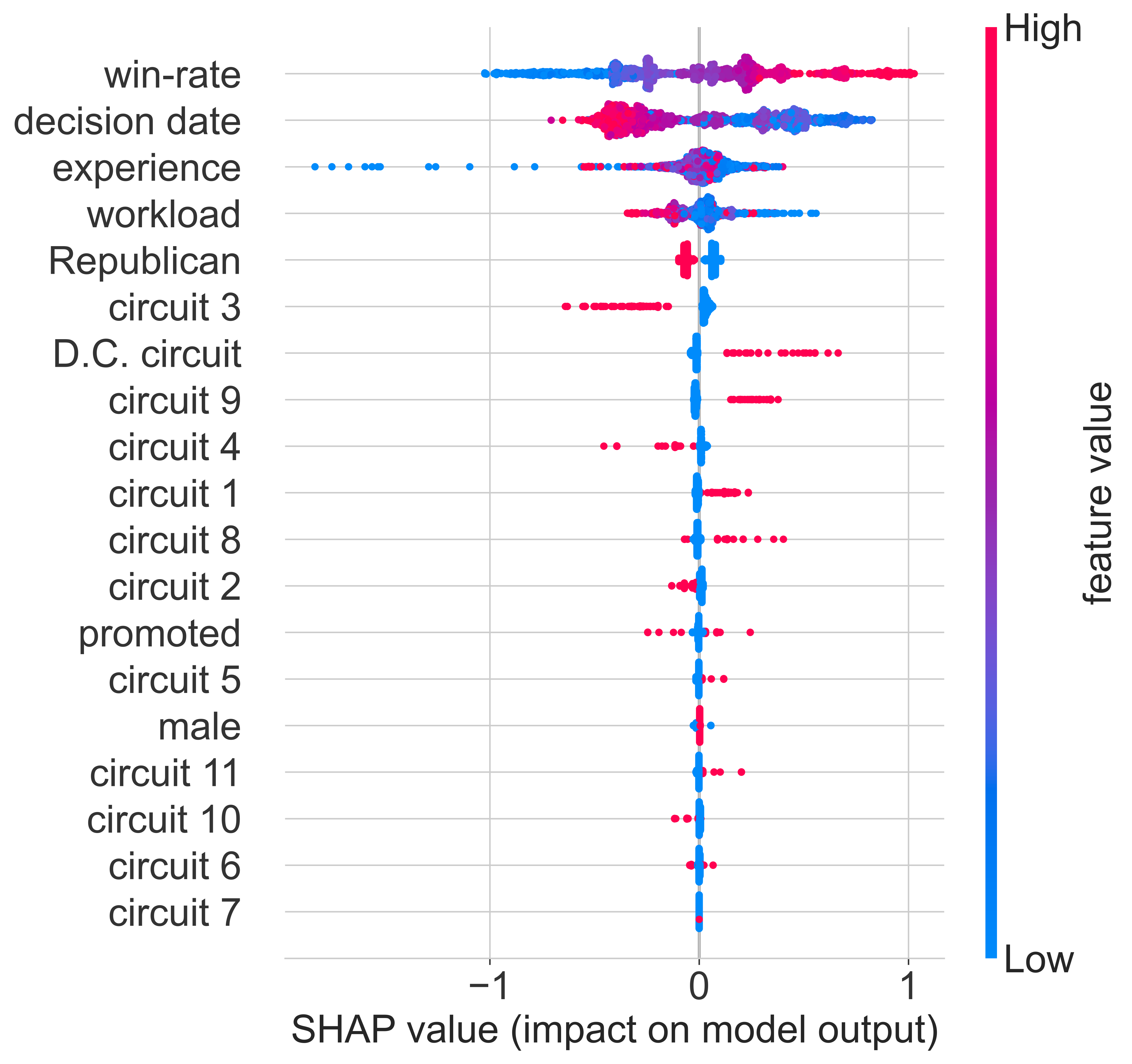}
    \caption{\textbf{Shapley feature importance for civil rights cases.} The feature importance highlights that past win-rate is the most important feature, with past high win rates indicating future high win rates. This suggests  a persistent effect of judicial idiosyncrasies. Similar plots for other case types are reported in the SI Appendix.}
    \label{fig:SHAP_overview}
\end{figure}

Average win rates vary significantly across case types, so we condition on the case type and train separate classifiers for each while controlling for circuits and decision dates.
To simplify interpretability across different predictions and case types, we down-sample both training and test data such that an equal number of cases are won and lost by the plaintiff in both samples.
A classifier that produces random guesses would thus exhibit an average accuracy of 50\%. 
As shown in Figure \ref{fig:biographic_feature_classification_results}, the overall prediction accuracy exceeds the 50\% baseline across all case types.
Performance may be magnified by leveraging the confidence scores assigned by the classifiers which ranges continuously from 0 (very confident that the plaintiff will lose the case) to 1 (very confident that the plaintiff will win the case). 
By binning these scores into quintiles, we find prediction accuracies for some case types (e.g. torts, civil rights, and prisoner petitions) exceed 70\% in the lowest and highest quintiles.
To further disambiguate the influence of different features on these relatively high prediction accuracies, we report the Shapley feature importance in Figure \ref{fig:SHAP_overview} for civil rights cases (see Methods for details on Shapley values and the SI Appendix for the other case types). 
The feature importances vary between case types, though the historical win rate, workload, and experience are consistently important judge-specific features.

We make several observations about these predictions:
First, 
we show that extraneous features unrelated to case facts or laws, such as the judge's workload, experience, and party affiliation, are predictive of case outcomes.
Given the random case assignment, this observation provides evidence that these extraneous factors affect judges' decisions.
The importance of these judge features varies between case types: 
they tend to be more predictive in torts, civil rights, and prisoner petition cases than in labor and contract disputes.
Second, 
we find that a judge's historical plaintiff win rate is persistent; 
judges with low plaintiff win rates are less likely to side with plaintiffs in the future, and vice-versa.
This suggests that the idiosyncrasies we observe may persist over judges' careers.
Third, 
and in line with previous research~\cite{lahav2018curious}, we find that for all case types other than prisoner petitions, plaintiff win rates have been declining and the decision date is consistently an important feature for our models.
Fourth, 
we find that the circuit in which a case is brought is predictive of its outcome.
This is unsurprising insofar as the precedent judges rely on varies between circuits.
However, it underscores that litigants' choice of forum can significantly impact litigation outcomes which is of particular relevance to national organizations, insurers, and litigation funders.

\subsection*{Prediction with Citation Information}
\label{sec:cite_prediction}

\begin{figure*}[!htb]
    \centering
    \includegraphics[width=\textwidth]{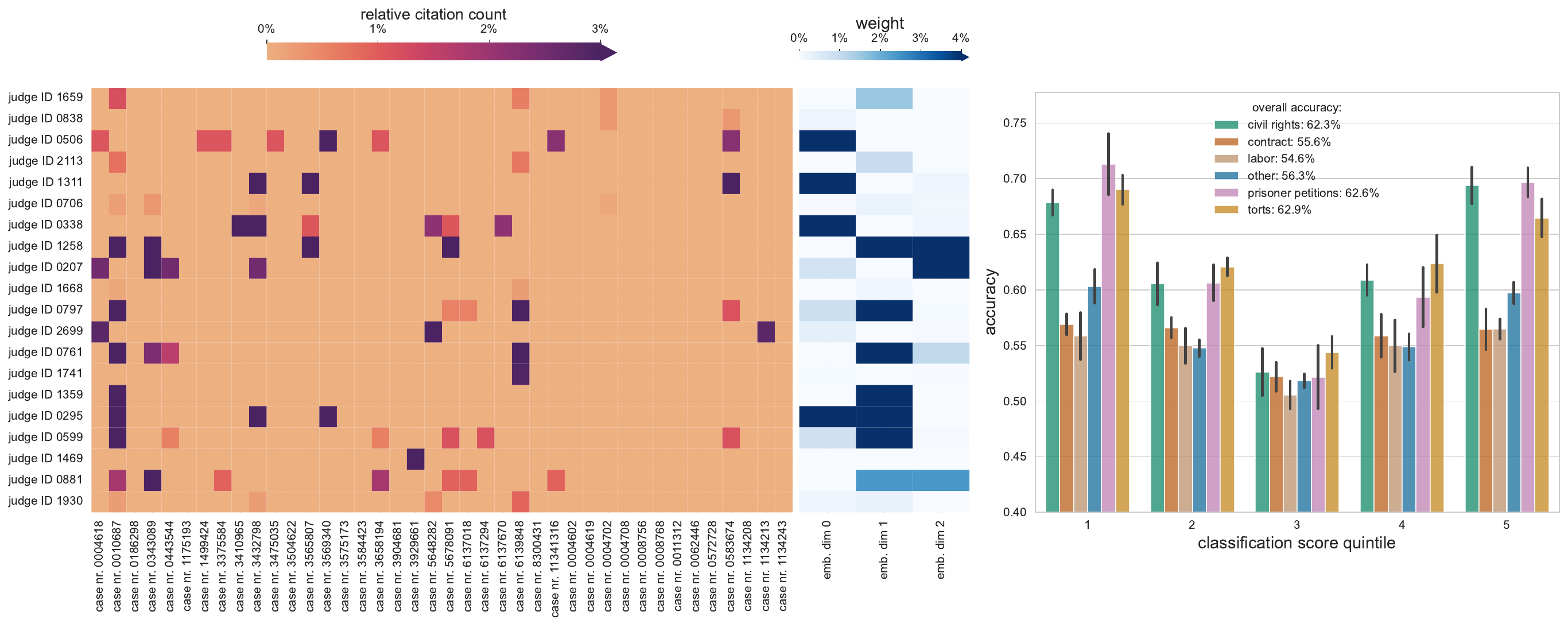}
    \caption{
            \textbf{Overview of predictions made with early-career citation embeddings.}
            (left)
            For a random selection of 20 judges, we show the normalized citation count of the 40 overall most commonly cited cases.
            We only count citations from the first 10\% cases of their career. 
            In practice, we consider for each of 2,394 judges the normalized count of the 2,403 most commonly cited cases. 
            (middle)
            NMF dimension reduction that compresses the 40 counts into 3-dimensional embeddings.
            In practice, we compress the counts of the 2,403 most frequently cited cases into 30 dimensional embeddings. 
            (right)
            Gradient boost classification accuracy based on these 30 dimensional embeddings, separated by confidence score quintile and case type.
            }
    \label{fig:citation_feature_classification_results}
\end{figure*}

The results above provide evidence that extraneous judge characteristics have an impact on judgments.
Information on judges is limited, and so it is not feasible to measure the effect of all plausible extraneous factors, in particular because many important variables related to judicial ideology are unobservable.
Instead, we take advantage of the fact that the U.S. is a common law jurisdiction, and judges must structure their arguments using citations to precedent.
There exists a large volume of precedent, and thus judges have some discretion in their citation decisions.
Citation decisions may therefore reflect judges' idiosyncrasies and, if so, past citations can be used to capture judicial idiosyncrasies.
Our methodology is further motivated by the observation that judges' historical win rates are highly predictive of future decisions, suggesting that idiosyncrasies are persistent.
We therefore aim to capture judicial ideologies by examining the citation decisions judges make at the start of their careers, in order to quantify preferences that form initially and persist over time.

For each judge, we generate an early-career citation record by aggregating all cases they cite in the first $10\%$ of cases assigned to them in our dataset.
There are approximately $1.7$ million published federal judicial opinions available to cite.
Due to this large number of cases, we only include citations to cases that were among the 500 most cited cases at end of a judge's early-career period (see Methods for details).

As shown in Figure \ref{fig:citation_feature_classification_results} (left),
these citation histories may be represented as a matrix $K$ where element $K_{i,j}$ represents the normalized count of judge $i$'s citation to opinion $j$. 
Since the most popular citations have changed over time, we consider a total of 2,403 cases across all judges.
To simplify training, and to avoid overfitting, we replace the 2,403-dimensional citation count features by lower-dimensional embeddings. 
Citation counts are non-negative and so we apply non-negative matrix factorization (NMF) to extract a low-dimensional representation of these citation features~\cite{Lee1999}.
Similar feature embeddings have previously been used 
to represent individuals' exploration patterns~\cite{yang2023identifying} 
or 
to find meta-genes from expression profiles~\cite{Brunet2004}.
See Methods for details and Figure \ref{fig:citation_feature_classification_results} (middle) for an example of a 3-dimensional NMF.
As shown in the SI Appendix, judgement prediction performance stabilizes at embedding dimensions of order $20$. 
To predict case outcomes, we thus fix the NMF dimensionality to $30$ and use these low-dimensional citation histories to train gradient boost classifiers to predict judges' decisions.
To avoid a look-ahead bias, we do not include the 10\% of cases used to generate citation histories as part of the classification data. 

We find that latent representations of early-career judicial behavior can be used to predict judgments with accuracies that match or exceed those of the biographic features used earlier (Figure \ref{fig:citation_feature_classification_results}, right).
This is despite the fact that the citation decisions used to construct the latent representations are \emph{completely unrelated} to judge's future decisions.
In line with our hypothesis, these citation features appear to capture judge's biographic characteristics (see SI Appendix for a regression of biographic variables against citation embeddings), providing further evidence that citation features capture judicial idiosyncrasies. 
It is especially notable that early-career citation histories appear to explain judge's promotions (Pseudo-R\textsuperscript{2} = 22.6\%), and we highlight this observation as an interesting area for future work.
Taken together, our findings support the conclusion that judges' idiosyncrasies are captured in their citation decisions and that early-career citation records may be used to predict future judgments.
Our findings suggest that these idiosyncrasies develop either before judges' appointments or early in their judicial careers.
Overall, these findings provide further evidence that judicial decisions in U.S. District Courts are affected by extraneous factors.

\begin{figure}[!htb]
    \centering
    \includegraphics[width=0.5\textwidth]{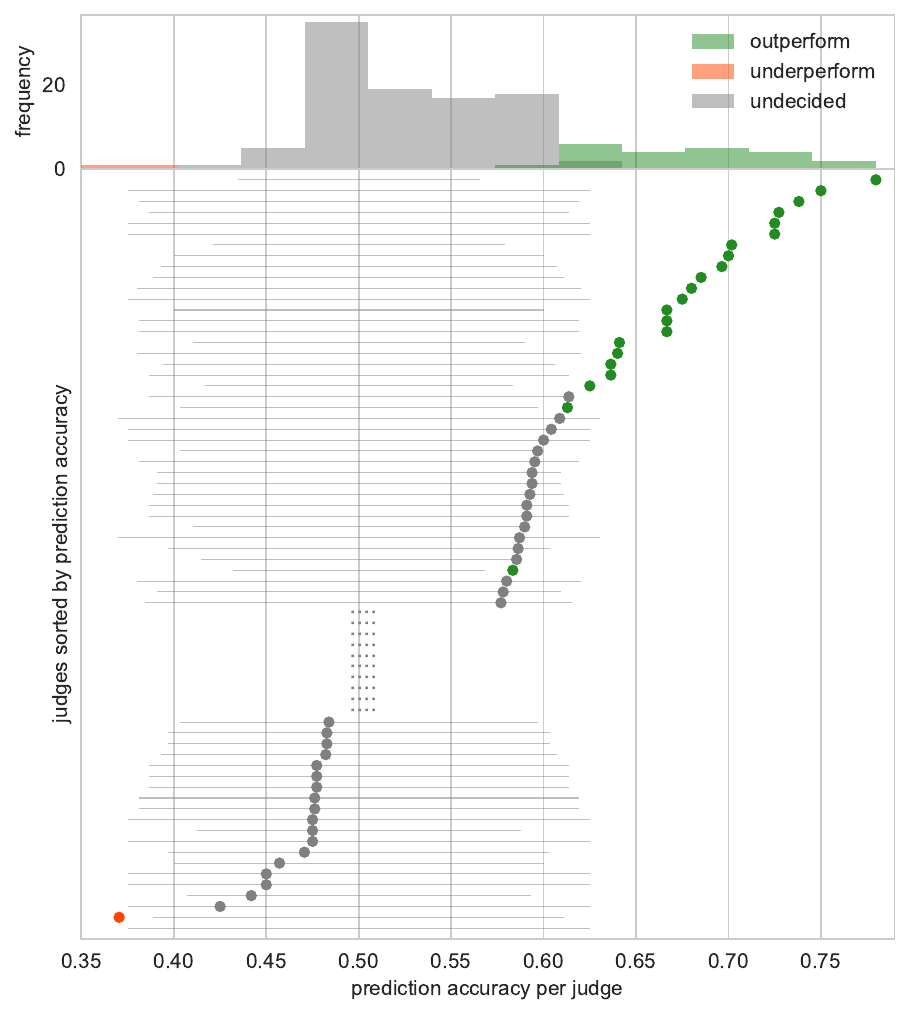}
    \caption{
    \textbf{Significance of aggregated predictions over judges' careers.}
    For each judge we balance the out of sample data such that half the cases are won by the plaintiff. 
    We only consider judges for which we have at least 50 out of sample cases.
    For each judge, we analyze the accuracy of predicting case outcomes.
    Our null hypothesis is that predictability is binomially distributed with $p=0.5$, i.e. there is no excess predictability.
    (top)
    Distribution of win rates, colored by whether the predictability is below (red), within (gray) or above (green) the 90\% confidence interval. 
    (bottom)
    Judges sorted by prediction accuracy along with their respective 90\% confidence intervals (gray bar). 
    Green (red) dots are significantly over- (under-) performing beyond the 90\% confidence interval of what is expected from random guessing.
    }
    \label{fig:predictability_per_judge}
\end{figure}

While these results may be surprising to many, at least some legal philosophers accept that judicial ideology may be used to resolve ambiguous cases (see e.g.,~\cite{coleman1993determinacy, dworkin1974hard}).
However, as discussed earlier, it is also generally accepted that these cases will be rare~\cite{howard2014courts, posner2010some, cardozo1924growth}.
We now examine whether some judges seem to \emph{systematically} lean on their ideology.
To this end, we analyze the decisions made by 224 judges for which we have at least $40$ balanced decisions in the test set (Figure \ref{fig:predictability_per_judge} and Methods).
For each judge, we determine the prediction accuracy of our NMF citation-based classifier.
A prediction accuracy that is significantly above 50\% indicates that that judge's cases can be confidently predicted by the judge's early-career citation history.
We test this by comparing the prediction accuracy for each judge against the 90\% confidence interval corresponding to the null hypothesis that the prediction is based on random guessing (binomial distribution with $p=0.5$).
If the null holds true, we expect roughly five out of 100 cases to be above or below the 90\% confidence interval, respectively.
We find that the model out-performs the baseline far more often than under-performing: for 22 judges we obtain predictions that are significantly above the 90\% confidence interval while only two judge's predictions are below. 
Even after applying a Benjamini-Hochberg correction to account for inflated false positives due to multiple testing (see Methods) we find that 6-8\% of judges remain positively biased.
This implies that around 50 of the current U.S. District Court judges may systematically rely on extraneous factors in reaching their decisions in civil disputes.

To further explore how citation histories capture judicial idiosyncrasies, we repeat the prediction task using raw citation counts rather than their low-dimensional representations.
This allows us to identify case citations with high Shapley feature importance to qualitatively investigate how different citations are associated with plaintiff friendliness (SI Appendix). 
The two cases with the highest Shapley feature importance for civil rights cases are \textit{Anderson v. Liberty Lobby (1986)} (Case ID: 6207800) and \textit{Celotex Corp. v. Catrett (1986)} (Case ID: 6206897).
We find that judges who disproportionately cite to these cases are associated with low plaintiff win rates, and vice-versa.
These Supreme Court opinions both concern summary judgment motions, that can be used by both defendants and plaintiffs to end a case early.
Legal scholars argue that \textit{Anderson} and \textit{Celotex}, tend to help defendants by making it easier for them to win a summary judgment motion~\cite{steinman2006irrepressible}.
By contrast, for \textit{Albemarle Paper Co. v. Moody (1975)} (Case ID: 9485) the Shapley value suggest that a high number of citation corresponds with high win rates.
Indeed \textit{Albemarle} tends to favor plaintiffs by making it easier for employees who have been discriminated against at work to demand back pay from their employer.
Judges have discretion over which cases they choose to cite and so their citation decisions provide insight into their reasoning. 
Our qualitative findings provide additional evidence that judges' citation behavior captures their idiosyncrasies. 

\section*{Discussion}
\label{sec:Discussion}

Judicial impartiality has been a central pillar of legal systems for millennia.
A range of safeguards exists to ensure that judges are not affected by extraneous factors when making their decisions.
In U.S. Federal Courts, these mechanisms include life-time appointments, judicial ethics rules, and random assignment of cases to judges.
Despite its importance, the large-scale quantitative study of judicial impartiality has been challenging due to limited data availability.
As a result, prior work on this topic has focused on higher court decisions or on a small number of lower courts.
While high courts, like the U.S. Supreme Court, have far-reaching influence, lower courts make routine decisions on many more cases and most of these decisions are not appealed to a higher court.
Much of the prior work on impartiality is focused on criminal law, which is by its nature highly consequential since criminal defendants face life-altering prison sentences.
However, impartiality in civil law, which governs relationships between people and organizations through rules on topics like contracts, employment, and negligence, also impacts a large fraction of society and has been studied less.
In an attempt to focus on judicial idiosyncrasies that affect every-day life, we thus focus on civil law decisions in the U.S. District Courts.
We leverage a novel dataset of U.S. District Court judgments to conduct a large-scale analysis of judicial impartiality.
We confirm that cases in these courts are assigned randomly, observe that a sizable fraction of judges significantly deviates from the expected win rate distribution, and use prediction methods to find evidence that extraneous factors affect judgments.

Numerous prior studies have found that judges are affected by a range of extraneous factors that are unrelated to case facts or laws 
including religious holidays~\cite{mehmood2023ramadan}, 
defendant characteristics~\cite{depew2017judges, alesina2014test, shayo2011judicial}, 
and food breaks~\cite{danziger2011extraneous}. 
We extend this prior work to 112,312 civil law cases heard in U.S. District Courts to measure the general effect of judges' idiosyncrasies on their judgments.
We initially use our data to confirm random case assignment.
This mechanism is crucial for impartial judging: if the assignment of cases were affected by the parties or by the judge this would raise an obvious potential for partiality.
Random assignment is also important for quantitative analysis because it eliminates the possibility that observed tendencies are the result of case assignment.
For almost $40\%$ of judges in our data, the career-averaged win rate significantly deviates from the expected baseline.
We find deviations on both sides of the win rate distribution with one judge ruling in the plaintiffs' favor in 107 out of 158 cases (68\%) while another sided with plaintiffs in 20 out of 330 cases (6\%).
The large number of outliers we identify suggests that many U.S. District Courts are influenced by extraneous factors.
We investigate the observed deviation by using judges' biographic characteristics such as their historical win rates, experience, gender, and party affiliation to predict their decisions.
We find that using just a small number of extraneous factors allows judges' decisions to be predicted, with accuracies exceeding $70\%$ for high-confidence predictions.
In principle, none of the biographic factors that are used in these predictions should have an effect on judges' decisions, and the predictability we demonstrate casts doubt on judicial impartiality at a large scale. 
Some of these preferences may be expected, after all, U.S. political parties expend significant resources identifying judges aligned with their ideologies.
However, we find that less obvious factors, such as historical win rate, workload, and experience, frequently appear to be more important predictors than party affiliation.

It is not possible to measure all factors that may potentially affect judges.
Instead, we take advantage of the fact that U.S. judges are operating in a common law regime where they build their arguments on citations to precedent.
Judges have tremendous discretion in choosing what precisely to cite and so examining judges' citation records provides insights into their ideology and idiosyncrasies.
We find that the citation decisions judges make early in their careers can be used to predict future case outcomes with a similar accuracy as biographic features. 
This suggests that citations capture judges' preferences and that judges form their idiosyncrasies early in their careers.
We thus show that low-dimensions embeddings of citation histories provide valuable latent representations that are sufficient to capture idiosyncrasies. 

Legal scholars broadly accept that some cases will be ambiguous and thus require a judge to exercise some discretion.
While there are diverging perspectives on what a judge should do in these instances, scholars broadly agree that such ``hard'' cases are relatively rare.
Coupled with random case assignment, we would thus expect that it is not be possible to systematically predict an impartial judge's decisions based on extraneous factors.
However, we find that for 6-8\% of judges, we are able to make statistically significant predictions across all their decisions, suggesting that a small number of judges systematically side with or against plaintiffs and that these judges may be relying on extraneous factors even when deciding ``easy'' cases.

Judicial impartiality plays an important societal role.
From a reform perspective, our finding that idiosyncrasies emerge early in judges' careers suggests that careful screening of judges prior to their appointment may help to promote impartiality.
The approaches we leverage in this study -- augmenting legal data with NLP and capturing judicial ideologies via citation records -- are generalizable beyond the U.S. legal context.
Over a third of the global population lives in a common law legal system, and we expect that our approaches could be applied in any such system to surface judicial idiosyncrasies.
While we focus on attitudes toward plaintiffs, our approach could readily be generalized to study judicial idiosyncrasies related to race, gender, religion, or socioeconomic status. 
Our dataset may also be used to study other aspects of litigation, such as law firm performance~\cite{mojon2024addressing}.
While it is very difficult to study judges through random control trials, citation histories appear to provide a window into judges' reasoning.
However, citation decisions are also likely to be confounded by a judge's court and other factors and future work is needed to better understand these relationships.
Some jurisdictions, most notably France, have restricted the use of quantitative methods to predict judicial behavior using personal data (this further motivates our use of citation decisions which are part of a judge's professional conduct).
Generally, these types of analyses can give wealthy litigants an edge or undermine trust in the judicial system by surfacing preferences, they can also be used as evidence for judicial reform or expand access to justice.
Moreover, the existence of judicial idiosyncrasies is undoubtedly known to practicing litigators, who use various methods to steer their clients towards the most favorable judges~\cite{kahan2021judge}.
While our analysis of random case assignment suggests that this is relatively uncommon, and the Judicial Conference of the United States, has been implementing rules to curb judge shopping, this creates an advantage for litigants with access to experienced lawyers.
Less well-resourced litigants may not have access to the same insights.
The quantitative approaches we demonstrate could thus level the playing field by allowing all litigants to identify optimal strategies or it may help bodies defining rules for the U.S. courts to more effectively mitigate judge shopping.

\section*{Conclusion}

Judicial impartiality has been recognized as a cornerstone of reliable court systems for thousands of years.
Despite its importance, quantitative study of impartiality has been challenging due to limited data availability.
We present the first large scale analysis of judicial reliance on extraneous factors in the U.S. Federal Court System.
We overcome data availability issues by merging several publicly available databases and augmenting the merged dataset using state-of-the-art natural language processing techniques to obtain a dataset of $112,312$ U.S. District Court decisions since 1880.
We confirm that cases are randomly assigned to judges, but despite this find that almost 40\% of judges' career-averaged win rate departs from the expected baseline.
We present evidence of judicial idiosyncrasies by demonstrating that that biographical features--like a judge's gender, party affiliation, and workload--can be used to forecast judgments with high-confidence prediction accuracies exceeding $65\%$ on a balanced test sample for certain case types.
We further show that judges' citation decisions capture their preferences and that early-career citation histories may be used to predict future judgments with similar accuracies as biographic features.
We find that these citation records are significant predictors for the aggregated decisions for roughly 6-8\% of U.S. District Court judges, which corresponds to around 50 of the currently active District Court judges.
Taken together, our findings highlight that a small but significant group of U.S. judges systematically relies on extraneous factors and that theses idiosyncrasies develop early in judges' careers and tend to persist.
A third of the global population lives in a common law jurisdiction and the methodology we present can be generalized to quantify partiality in any such system without requiring access to judge's biographic data.
Our methodology may also be generalized beyond plaintiff-friendliness to capture idiosyncrasies related to other litigant characteristics including race, religion, gender, or socioeconomic status.
Our findings can be used to promote reform efforts to increase judicial impartiality by providing quantitative evidence of extraneous influences and by highlighting that careful vetting of judges prior to their appointment is may be an effective remedy for over reliance on extraneous factors.

\section*{Data Availability Statement}

The data used in this work is freely available from the Case Law Access Project (CAP) and the Federal Judicial Center (FJC). 
We use the full federal judicial opinion dataset from CAP as well as the citations graph provided by CAP, both of which are available at \url{https://case.law/}. 
The judge demographics data is obtained from \url{https://www.fjc.gov/history/judges}. 
The FJC's case data may be downloaded from \url{https://www.fjc.gov/research/idb}.
We make our data available for reviewers and will release the full dataset upon publication.

\section*{Code Availability Statement}
The code to reproduce all analysis from this paper can be found via GitHub at \url{https://github.com/lerasc/judge_predictability}.

\section*{Author Contributions}
R.M. and S.C.L. contributed equally in the conceptualisation, methodology, analysis, and writing.

\section*{Competing Interests}
The authors declare no competing interests.

\section*{Methods}
\label{sec:Methods}

\subsection*{Data}
\label{method:data}

The data for this project was obtained by merging four datasets.
The first is the Integrated Database (IDB) provided by the Federal Judicial Center (FJC). 
It contains tabular data on federal civil litigation since 1970, including when a case was filed, in which judicial circuit it was filed, the plaintiff and defendant names, and the case outcome. 
The U.S. Federal Court System features three hierarchical levels: 94 District Courts, 12 regional Appellate Courts, and the Supreme Court. 
This study focuses on District Courts which tend to make routine legal decisions that are likely to impact everyday life and which remain understudied.
The 94 U.S. District Courts are organized into 12 geographic circuits.
In the IDB, lawsuit outcomes have three possible labels: the plaintiff wins, the defendant wins, or ``unknown''. 
We discard unknown outcomes, which likely correspond to private settlements, because they cannot be matched to the other datasets.
The IDB assigns a ``Nature of Suit'' code to each lawsuit, and we aggregate these granular codes into the following categories using a mapping published by the Public Access to Court Electronic Records service: 
civil rights, contract, prisoner petitions, torts, labor, and ``other''. 
There are a few more niche case types (such as immigration, bankruptcy, and social security) which account for a minor fraction of cases and are thus assigned to the case type ``other''.
Overall, the IDB contains roughly 2.5 million District Court cases for which it reports tabular data including case outcomes and case types.
Crucially, however, the IDB does not contain information about the presiding judge.

The second dataset is the Case Law Access Project (CAP) which contains judicial opinion texts scanned and digitized from court reporters over the last 360 years.
The CAP data includes case names (normally of the form ``party\_1 v. party\_2''), the court names, and raw opinion texts in which presiding judge(s) explain the case facts, applicable laws and justify their decision.
The CAP metadata includes judge names, usually in the form of last names e.g. ``POSNER, Circuit Judge''.
While the CAP data does not contain explicit information about case outcomes and case types, that information is implicitly part of the judge's opinion. 
Each case is assigned a unique case number, which we rely on to uniquely identify cases.
The CAP database contains both civil and criminal cases.
For our analysis, we discard all criminal cases which we identify as cases where one of the parties is the \textit{United States} (there are some civil cases involving the U.S., but these are relatively uncommon).
We further restrict ourselves to District Court cases by filtering the court names.
Following these steps, we are left with 302,986 civil District Court cases for which judicial opinion is available and the identity of the presiding judge is known.

The third data-set is a citation network, which is also obtained from the CAP and which shows citations between the CAP opinions. 
For each opinion, the citation network contains a list of all opinions that it cites.
We rely on this network to obtain early-career citation histories for judges.

The fourth dataset is the Judges Dataset maintained by the FJC which contains judges' biographical information including gender, appointment and termination dates, the identity of the U.S. President who appointed them, and information on promotions.
Judges are identified by a unique number, as well as their first and last name, birth year, death year (if applicable), and nomination date. 
By contrast, the CAP data only identifies judges by their last names, which are not necessarily unique.
We disambiguate judges with identical last names via their initials, the court, and the timing of their appointment.
We then replace the name of each judge by the unique FJC identifier (as shown in Figure \ref{fig:citation_feature_classification_results}).

\subsection*{Mapping datasets}

One of the main practical challenges for this study is mapping judges' decisions in form of raw opinion texts (CAP) to tabular case outcomes and case types (IDB).
The easiest way to map these would be via \textit{case docket numbers}, which are, however, not always available and not always unique. 
We thus map between CAP and IDB data based on the docket number where possible, and else via plaintiff and defendant names, court names, and dates. 
Unfortunately, due to OCR errors and other inconsistencies, we only obtain satisfactory mappings for 31,222 cases.
To further augment the data, we thus leverage recent advances in natural language processing. 
For the 31,222 mapped cases we have information about the case outcome (the plaintiff either won (1) or lost (0) the case), as well as information about one of six case-types from the IDB. 
For these cases, we have rich textual information about the details of the case from the CAP opinions and we train two transformer-based classifiers on the 31,222 successfully mapped cases. 
These classifiers assign a case outcome (binary classification task) and case type (classification task with 6 labels) based on the opinion text.
We then apply the trained classifiers to the opinion texts of all 302,986 civil  District Court cases in the CAP database to obtain out-of-sample classifications for case outcome and case type.
To minimize the possibility of incorrect labels, we subsequently discard all cases for which the softmax output of the final layer is too close to 0.5.
For the case outcome classifier, we only consider cases with a final softmax output below 0.1 or above 0.9. 
For the case type classifier, we only consider cases where the largest of the 6 softmax outputs is above 0.5. 
Considering out-of-sample predictions with softmax outputs above those thresholds yields prediction accuracies above 95\%, as determined on a separate validation dataset (see SI Appendix).
Based on these cut-offs we obtain a final dataset of 112,312 cases which form the basis of our analysis. 

The NLP classifier we use is the \textit{Longformer}, a pre-trained transformer model developed for long documents~\cite{beltagy2020longformer}.
The Longformer is able to process 4096 tokens, rather than the 512 tokens that can be handled by common transformer models like BERT.
This attribute makes the Longformer well-suited for judicial opinions which tend to be long documents.
Even so, some opinions exceed the maximum input length and for these opinions we use only the final 4096 tokens since the end of judicial opinions tends to contain a summary of the decision.
We fine-tune the pre-trained Longformer on each classification task (outcome and case type) for 5 epochs with cross-entropy loss. 

\subsection*{Random assignment of cases to judges}
\label{method:random_assignment}

U.S. District Courts explicitly state that cases are assigned to judges at random.
We examine this by conducting a large-scale test of random assignment based on the observed case type of assigned cases.
Our analysis includes all 2,394 judges that issued at least ten judgments in a given circuit during a given decade.
For each circuit and decade, we determine the rate of occurrence (base-rate) for each case type by dividing the number of cases of that type by the total number of cases.
We only consider case types, circuits and decades with at least 100 cases to make sure that the base-rate is representative. 
For each judge (in that circuit and decade), we then calculate the fraction of cases of each type relative to the total number of cases that the judge presided over.
For each case type and judge, if there is no bias in case assignment, we expect the number of cases to be binomially distributed with frequency equal to the base-rate. 
We thus test the null hypothesis that the case assignments are sampled from a binomial distribution by using a two-sided binomial test, with the case type base-rates for a given circuit and decade as a reference.
The resulting p-value can be interpreted as the probability of the observed assignment, assuming the the null hypothesis (random assignment) holds true. 
For example, consider a situation where a judge presides over 100 cases, of which 70 are civil rights cases. 
If the base-rate for civil rights cases is 40\%, the associated p-value is of order $10^{-8}$, which makes it very unlikely that the case assignment was random. 
By contrast, if the number of civil rights cases is equal to 45, the associated p-value is 31\%, which does not provide strong evidence against random assignment.

Grouping cases by case type, decade and judge leaves us with a total of 18,355 datasets, each on which we can test the null hypothesis of random assignment.
In principle, we could now analyze each of the 18,355 p-values.
For instance, a typical approach is to reject the null hypothesis if the p-value is below 5\%. 
However, given this high number of tests, the likelihood of encountering false positives (type I errors) increases (see SI Appendix). 
To address this, corrections for multiple comparisons are necessary. 
The Benjamini-Yekutieli procedure is particularly suitable in this context \cite{benjamini2001control}. 
It allows for the identification of significant findings while accounting for dependency structures. 
Indeed, our p-values are not perfectly independent since we had calculated the baseline probability as the average case frequency across circuit and decade.
After applying the Benjamini-Yekutieli correction, we find that a total of 0.6\% of all p-values fall below the 5\% threshold.
Overall, this suggests that the case assignment is largely random in the sense that for 99.3\% of our datasets we cannot reject the null hypothesis that case types are randomly assigned.

In the SI Appendix, we provide additional graphical evidence for random case assignment. 
If the null holds true, the $p$-values ought to be uniformly distributed.
We test this by plotting empirical quantiles against theoretical ones and measuring their Pearson correlation. 
We find empirical support for random case assignment with correlations of order 80\%-95\%. 

It is possible that case assignment biases may manifest on more subtle grounds. 
To explore this further, we utilized the CAP dataset, which includes tabulated information about the names of the involved parties.
We categorized the cases into four distinct groups based on the nature of the parties involved: 
(i) cases involving the government as one of the parties,
(ii) cases where both parties are companies,
(iii) cases where both parties are private individuals, and
(iv) cases with a mix of party types.
To systematically extract this information across all 112,312 cases, 
we employ OpenAI’s GPT-4 model, prompting it to assign each case to one of the four categories (see SI Appendix for details).

We then apply the same methodology used in our initial case-type analysis. 
Specifically, we group all cases by jurisdiction, decade, and judge, resulting in 15,133 distinct datasets. 
For each dataset, we tested the null hypothesis of random case assignment, comparing each group to the baseline average across all cases. 
After applying the Benjamini-Yekutieli (BY) correction, only 0.4\% of hypotheses are rejected, 
further supporting the inability to reject the null hypothesis of random case assignment.

\subsection*{Features}

We use three classes of features: control variables, bibliographical judge information, and citation histories.
(The distribution of each feature is shown in the SI Appendix.)
To avoid a look-ahead bias, all the features are calculated using only information that would have been available at least 60 days before the case decision date.

The control variables include the case type, the date the case was decided between 1880 and 2018, and the twelve geographical circuits, which we represent as 12 binary features via 1-hot encoding.

We take the following steps to prepare the bibliographical judge information.
We infer the judges' party via the party of the appointing president.
Judges that serve on multiple courts are counted as ``promoted'' from the time of their promotion onwards. 
The judge's gender is taken from the FJC's Judges dataset.
We calculate the judges' historical win rate as the average plaintiff win rate across all cases that were presided over by the judge since the beginning of their career until 60 days prior to deciding on the case. 
The judges' workload is calculated as the average number of cases per year, averaged over each year prior to a given case. 
Finally, the experience is calculated as the time (in years) between a judge's appointment date and 60 days prior to the case outcome.

We obtain the citation histories for each judge by following the following steps:
For each judge, we define the early-career window as the span of years in which the judge adjudicates the first 10\% of cases. 
For example, if a judge is appointed in 1970, presides over 100 total cases and decides the tenth case in 1980, then the early-career window would span from 1970 to 1980. 
For each judge, we identify the 500 most cited cases across all judges in the year prior to the end of the relevant early-career window (in the prior example that would be 1979) and determine how often the judge cited to these cases during the early-career window. 
We then normalize the citation counts such that they sum up to one. 
We repeat this procedure for each judge and unify the result in a single citation history matrix.
The most popular citations have changed over time and so there are 2,403 unique citations in the matrix and we impute 0 for any cases that a judge did not cite.
This results in a feature matrix $C \in \mathbb{R}_{\geqslant 0}^{m \times n}$ where $m$ is the number of judges and $n$ the number of cases. 
We show a reduced version of this matrix in Figure \ref{fig:citation_feature_classification_results} (left) for $m=20$ and $n=40$. 
These $n$-dimensional feature vectors are subsequently compressed via non-negative matrix factorization (NMF) into $k \ll n$ dimensions.
The goal of the NMF is to approximate $C \approx WH$ where $W \in \mathbb{R}_{\geqslant 0}^{m \times k}$ and $H \in \mathbb{R}_{\geqslant 0}^{k \times n}$. 
The matrix $H$ thus stores the latent citation patterns whereas the low-dimensional feature matrix $W$ contains the weights of those latent patterns for each judge. 
Practically, we use the implementation from Python's scikit-learn package which minimizes the square of the Frobenius norm $\left | \left| C - W H \right| \right|^2$
along with $L1$- and $L2$-regularization terms for both $W$ and $H$. 
Figure \ref{fig:citation_feature_classification_results} (middle) shows an example of the matrix $W$ for $k=3$. 
In practice, we have used $k=30$, but convergence is already reached for values as low as $k \approx 20$ (SI Appendix).

\subsection*{Case Outcome Classification}

Given the features described above, we perform binary classification to predict whether the plaintiff wins (1) or loses (0) a case. 
We use XGboost's gradient boost classifier for this task, which is a non-linear, tree-based model.
As we show in the SI Appendix, similar, albeit slightly less accurate results are obtained for the random forest, multilayer perceptron and logistic regression with Ridge L2-regularization. 
We use $3$-fold cross-validation to tune the model's hyper-parameters and optimize for prediction accuracy. 

We train a total of 12 classifiers: 
For each of the 6 case types, we train 
one classifier with biographic judge features
and
one classifier with citation features.
In all models we include decision date and circuits as control variables.

For each of the 12 classifiers, we apply a $75$\%-$25$\%  train-test split.
We subsequently balance both the train- and test-data so that the number of wins and losses are equal in both to simplify the interpretation and comparison of results across case types.
All features are standardized by subtracting the mean and dividing by the standard deviation, as determined on the training data. 
For the citation classifiers only, we discard the initial 10\% of cases assigned to a judge to avoid a lookahead bias since these were used to calculate the early-career citation history.

We assess the success of prediction on the test set via accuracy score.
Rather than assigning a predicted outcome to each case, the classifier outputs a confidence score ranging from 0 to 1.
We label cases with predicted probabilities above/below $50\%$ as won/lost by the plaintiff. 
Further, we bin cases by their predicted probability into five buckets 
--
$
[0-0.2), ~[0.2-0.4), ~[0.4-0.6), ~[0.6-0.8), ~[0.8-1.0)
$
--
to assess the accuracy as a function of confidence 
(Figures 
\ref{fig:biographic_feature_classification_results}
and 
\ref{fig:citation_feature_classification_results}).
As anticipated, the accuracy is typically higher the more the score deviates from $0.5$. 
This is particularly relevant for those interested in predicting the likely outcome of potential lawsuits.
For instance, a third-party funder may only be willing to invest in cases with a predicted score greater than some threshold \cite{Lera2022}.
As a result of the data balancing, a naive classifier would result in an accuracy of $50\%$, which is significantly exceeded in the outer bins across all case types. 
To minimize the effect of statistical fluctuations from the down-sampling, we bootstrap the out-of-sample test data and report average accuracy and the standard deviation as error bars. 

To assess these predictions across a judge's aggregated decisions, we generate predictions in the same manner as above, but this time grouped by judge.
For each judge, we balance the out of sample data such that an equal number of cases are won and lost by the plaintiff, respectively. 
We subsequently focus on 224 judges for which at least 50 out-of-sample text cases are available (qualitatively similar results are obtained with a slightly higher or lower cutoff).
Our null hypothesis is that for each judge predictability is binomially distributed with $p=0.5$, i.e. there is no excess predictability above 50\% beyond what is expected by chance. 
For each judge, we thus calculate the 90\% confidence interval corresponding to the null hypothesis based on the number of cases available (Figure \ref{fig:predictability_per_judge}).
If the average prediction accuracy falls outside of these confidence intervals, then this is evidence that citation features predict this judge's decisions.
We find that predictions for 26\% of judges fall significantly outside of the confidence intervals.
However, since we are repeatedly testing the same hull hypothesis, we follow the procedure elaborated on above, and apply a Benjamini-Hochberg correction factor to the 224 p-values.
This results in a fraction of 5-10\% of judges which deviating significantly from the null hypothesis
(see SI Appendix for details on the correction methods). 

Recently, Shapley values have gained prominence in the field of machine learning as a technique to mitigate the opacity of model predictions. 
By assigning each feature in a dataset a fair and quantifiable share of the prediction's outcome, Shapley values provide insights into the relative importance and contribution of individual features~\cite{Chen2023}. 
Figure \ref{fig:SHAP_overview} (and more figures in the SI Appendix) shows global feature importances where a positive/negative Shapley value indicates a positive/negative contribution to the predicted probability that the plaintiff will win the case.
Each point represents one case, and its color represents the associated relative magnitude of the feature value.
Unlike the interpretation of regression coefficients from linear models, Shapley values can also uncover non-linear and non-monotonic feature interdependencies.

\section*{Acknowledgements}

We thank the following people for their generous feedback on this work: Morgan Frank, Matt Groh, Andreas Haupt, Ivy Luo, Takahiro Yabe, and Alexandre Mojon. 

\balance
\bibliographystyle{naturemag}  
\bibliography{bibliography} 

\newpage
\onecolumn
\appendix
\setcounter{figure}{0}   
\renewcommand\thefigure{SI \arabic{figure}}    
\renewcommand{\thetable}{SI \arabic{table}}  
\label{SI}

\section{Transformer Based Case Labeling}
\label{SI:transformer_labeling}

The core dataset that forms the foundation of our study is the Harvard Case Law Access Project (CAP) dataset, which includes judicial opinion texts detailing how a judge decided a case. 
We extract two important features from these opinions for two primary reasons:
first, whether the plaintiff won or lost the case, and second, to identify the case type. 
To systematically extract this information, we finetuned two Longformer models \cite{beltagy2020longformer}: 
one for predicting case outcomes and another for identifying case types. 
Below, we provide a more detailed explanation of how these Longformer predictions were incorporated into our final dataset.

It is important to note that, apart from the technical challenges associated with natural language processing, this task is relatively straightforward from an information-theoretic perspective. 
The information regarding both case type and case outcome is explicitly embedded in the judges’ opinions. 
This contrasts with the more complex task - which forms the primary contribution of our paper -  of predicting case outcomes based on the characteristics of the judges.

\subsection{Case Outcome Classifications}

As elaborated in the methods section, we fine-tuned a transformer-based text classifier that maps a judge’s case opinion to a binary outcome: Plaintiff won (1) or lost (0) the case. 
Our training data comprises 31,222 cases, of which we reserve 10\% for testing. 
When given a judge’s opinion, the trained classifier does not predict a binary outcome directly but instead produces a sigmoid output, $p$, ranging from 0 to 1, which is interpreted as the classifier’s confidence level. 
One could then classify each opinion with $p > 0.5$ as the plaintiff winning the case and as the plaintiff losing the case otherwise. 
However, it is reasonable to assume that the decision boundary around 0.5 contains a large number of false predictions. 
Therefore, we consider the ``classification confidence threshold'' $\tau \in (0, 0.5)$ and its effect on the quality of classifications. 
Specifically, at level $\tau$, we only consider the classification of cases where $| p - 0.5 | > \tau$.

\begin{figure*}[!htb]
    \centering
    \includegraphics[width=0.8\linewidth]{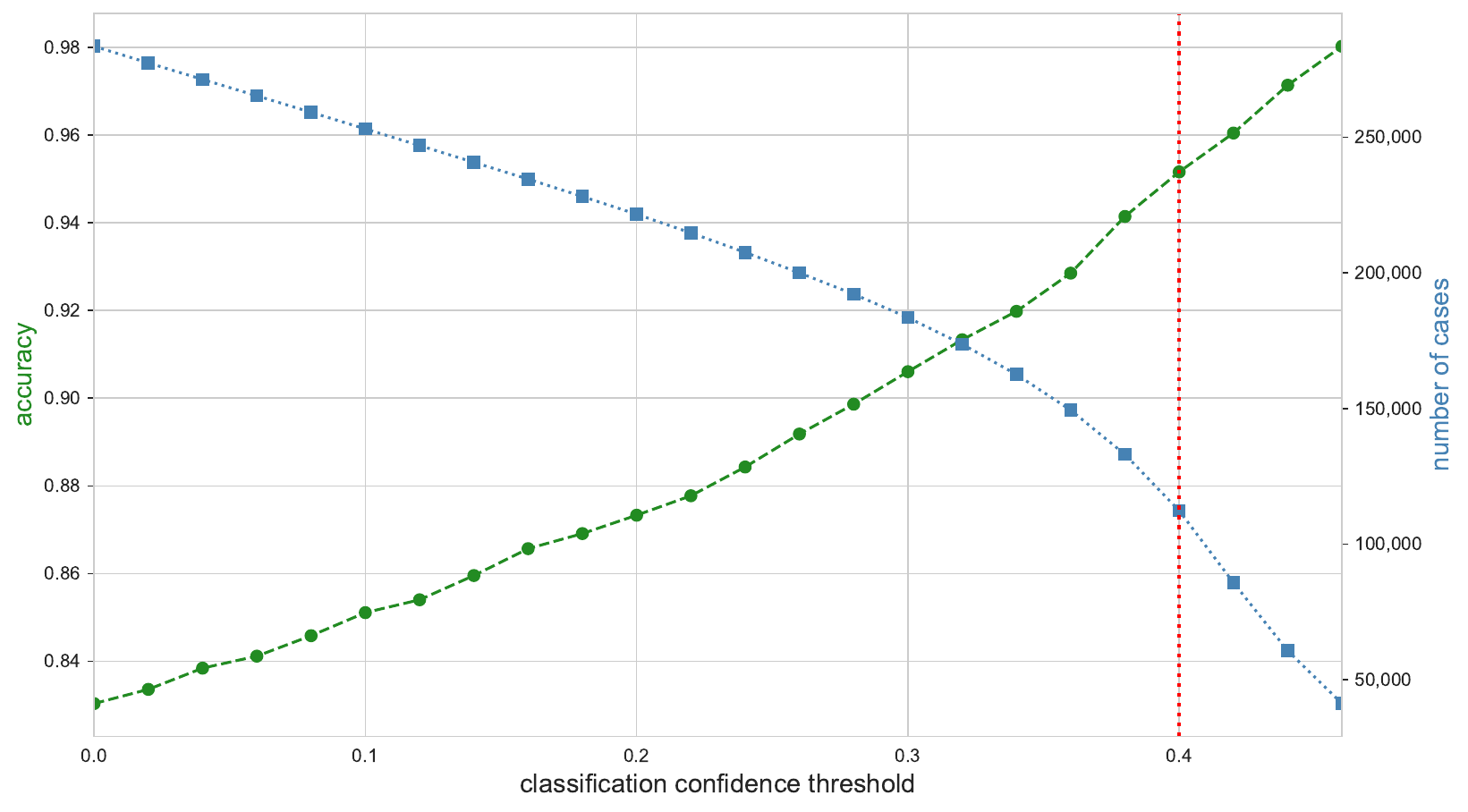}
    \caption{\textbf{Trade-off between data augmentation accuracy and dataset size.} Accuracy on test data (left y-axis) and number of available cases (right y-axis) as a function of the transformer classification confidence cut-off $\tau$.}
    \label{fig:transformer_classifier_sensitivity}
\end{figure*}

Figure \ref{fig:transformer_classifier_sensitivity} shows the prediction accuracy on the 10\% test data as a function of $\tau$ on the left y-axis. 
On the right y-axis, we show the associated number of out-of-sample cases available at that level of $\tau$. 
(Here out-of-sample refers to data for which we do not have a training label, see also summary in \ref{SI:transformer-summary}). 
While the transformer already yields respectable accuracy for low values of $\tau$, we opt for a trade-off between maximal data quality and the number of available cases.
We have thus decided to set a cut-off at $\tau = 0.4$ which corresponds to an outcome labeling accuracy of 95\% and leaves us with a total of 112,312 cases.
In other words, we consider only cases where the transformer’s final sigmoid output is either above 0.9 or below 0.1, discarding cases otherwise. 
However, as we show in \ref{SI:sensitivity_analysis}, our results remain qualitatively unchanged for lower values of $\tau$, albeit at slightly lower accuracies. 

\subsection{Case Type Classifications}

A similar analysis can be conducted for our second transformer, which has been trained to predict the case type. 
The transformer’s output is represented as a softmax distribution over the six possible case types. 
Denote these outputs by 
\( p_{\text{civil rights}}, p_{\text{contract}}, \dots, p_{\text{torts}} \). 
These outputs are normalized such that their sum equals one. 
If the transformer were to assign equal probabilities to each case type, 
it would hold that $p_{\text{case type}} = \frac{1}{6} \approx 16\%$ for each. 
To retain only cases where we are reasonably confident that the case type is predicted correctly, we filter the predictions by considering only those instances where the highest softmax output exceeds 50\%, and then assign the case to that corresponding case type.
This achieves an accuracy of 89\% on the validation data.

\subsection{Summary}
\label{SI:transformer-summary}

We begin with an initial set of 302,986 cases, for which we have the judge’s opinion from the CAP dataset.
Of these cases, we obtain labels for both case outcome and case type for 31,222 cases through mapping to the IDB dataset. 
We utilize this subset of 31,222 cases to train two transformer models: 
one to predict each of the six case types, and another to predict the case outcome (whether the plaintiff won or lost the case).

Once the case type transformer was trained, we applied out-of-sample predictions to the entire set of 302,986 cases, excluding those used in the training process. 
We retained only those cases where the highest softmax output for case type prediction exceeded 50\%, 
resulting in a total of 283,511 cases. 
We then applied the second transformer to these 283,511 cases to predict the case outcome. 
Based on the analysis summarized in Figure \ref{fig:transformer_classifier_sensitivity}, we selected a confidence threshold of 0.4, discarding all cases that were classified with lower certainty.
This left us with a total of 112,312 cases.

These 112,312 cases constitute the main dataset used throughout this study. However, as demonstrated in \ref{SI:sensitivity_analysis} below, qualitatively similar results hold for larger data subsets.

\clearpage
\newpage
\section{Testing Random Case Assignment}
\label{SI:random_assignment}

\subsection{A Primer on p-Value Adjustments}
\label{SI:p-value_primer}

Several of our result results depend on the testing a certain null hypothesis $H_0$ across multiple datasets. 
This requires careful adjustment of the p-values to avoid inflating Type I errors. 
This subsection provides an overview of three common correction methods. 

Assume that we have $K$ (independent) datasets. 
On each of the $K$ datasets we test the null hypothesis $H_0$.
Let us further assume that $H_0$ holds across all $K$ datasets. 
In that case, the p-values are uniformly distributed on the interval $[0,1]$. 
This can be seen as follows: 
If the null hypothesis \(H_0\) is true, the test statistic \(X\) follows the distribution expected under \(H_0\). 
Let \(F_0(x)\) denote the cumulative distribution function (CDF) of this null distribution, such that \(F_0(x) = P(X \leq x)\). 
The p-value is the probability of observing a test statistic as extreme or more extreme than the observed value under \(H_0\). 
For example, in a one-sided test, the p-value could be written as \(p = 1 - F_0(X)\) (for a right-tailed test) or \(p = F_0(X)\) (for a left-tailed test). 
A key result in probability theory is that if \(X\) is a random variable with CDF \(F_0\), then the transformed variable \(U = F_0(X)\) is uniformly distributed on the interval \([0, 1]\).
Therefore, under \(H_0\), the p-values \(p = 1 - F_0(X)\) or \(p = F_0(X)\) are uniformly distributed on \([0, 1]\).
Since p-values are uniformly distributed on \([0, 1]\) under the null hypothesis \(H_0\), a QQ-plot comparing the observed p-values to a theoretical uniform distribution provides a visual check for deviations from this expectation, indicating potential departures from \(H_0\). 
Examples of such plots are found in Figures \ref{fig:case_type_bias_examination} and \ref{fig:entity_label_bias_examination} below testing the null hypothesis for cases that were assigned to judges at the same base rate as the overall prevalence of such cases. 

While QQ-plots can reveal deviations from the expected uniform distribution of p-values, they do not control the increased risk of Type I errors (incorrectly rejecting $H_0$) in multiple hypothesis testing, making it essential to apply multiple-testing corrections. 
When performing a single hypothesis test at significance level \(\alpha\), there is a probability \(\alpha\) of making a Type I error. 
However, when conducting \(K\) independent tests, the probability of making at least one Type I error increases. 
The probability of not making a Type I error in a single test is \(1 - \alpha\), so the probability of not making a Type I error in all \(K\) tests is \((1 - \alpha)^K\). 
Therefore, the probability of making at least one Type I error across all \(K\) tests is \(1 - (1 - \alpha)^K\). 
For example, with \(\alpha = 0.05\), if \(K = 2\), this probability is approximately 9.75\%, and if \(K = 10\), it increases to about 40.13\%. 
This demonstrates the need for multiple testing corrections to control the overall error rate.

There are three relatively common adjustment procedures: 
The Bonferroni correction (BF), 
the Benjamini-Hochberg correction (BH)
and the Benjaminini-Yekutieli correction (BY) \cite{benjamini2001control}.

The Bonferroni correction controls the family-wise error rate (FWER), which is the probability of making at least one Type I error (false positive) across a set of multiple hypothesis tests. 
To control the FWER, the Bonferroni correction divides the significance level \(\alpha\) by the number of tests \(K\), setting a stricter threshold of \(\frac{\alpha}{K}\) for each individual test. 
This ensures that the overall probability of making any Type I error remains at or below the original significance level \(\alpha\), making it a conservative approach that minimizes the risk of false positives across multiple comparisons.

The Benjamini-Hochberg (BH) procedure is often considered better than the Bonferroni correction in scenarios involving multiple hypothesis testing because it controls the false discovery rate (FDR) rather than the family-wise error rate (FWER) \cite{benjamini2001control}. 
The BH method allows for a more balanced approach by controlling the proportion of false positives among the rejected hypotheses, rather than minimizing the probability of making any Type I error. 
This makes it less conservative than the Bonferroni correction, resulting in greater statistical power, meaning it has a higher likelihood of detecting true effects without sacrificing too much control over false positives.
The BH procedure works by first sorting the p-values in ascending order. 
For each p-value \(p_{(i)}\) with rank \(i\) among \(K\) total tests, the method compares \(p_{(i)}\) to an increasing threshold \(\frac{i}{K} \cdot \alpha\), where \(\alpha\) is the desired false discovery rate (FDR). 
The largest p-value that is less than or equal to its threshold is identified, and all hypotheses with p-values less than or equal to this value are rejected. 
By using an increasing threshold, the BH procedure controls the FDR, allowing for more rejections compared to the Bonferroni correction, while still maintaining a balance between detecting true positives and controlling the rate of false discoveries.

The Benjamini-Yekutieli (BY) method is an extension of the BH procedure that accounts for dependencies among the test statistics.
Like the BH method, the BY procedure starts by sorting the p-values in ascending order and comparing each p-value \(p_{(i)}\) to an increasing threshold. 
However, to handle dependencies, the BY method adjusts the threshold by a factor based on the harmonic series 
\(H(K) = \sum_{i=1}^{K} \frac{1}{i}\), 
where \(K\) is the total number of tests.
In the BH procedure, the comparison threshold for each p-value is \(\frac{i}{K} \cdot \alpha\), where \(i\) is the rank of the p-value and \(\alpha\) is the desired false discovery rate (FDR).
In the BY method, this threshold is further divided by \(H(K)\), so the new comparison threshold for each p-value becomes \(\frac{i}{K \cdot H(K)} \cdot \alpha\). 
The harmonic series \(H(K)\) grows logarithmically with \(K\), so as the number of tests increases, the thresholds for rejecting hypotheses become more conservative, reflecting the increased difficulty of controlling FDR under dependency.

After applying the BH or BY correction at the 5\% threshold, assuming all null hypotheses \(H_0\) are true, we expect most p-values to be above the 5\% level. 
However, due to random fluctuations, some p-values might still fall below the threshold, though far fewer than 5\%. 

To test random case assignment, 
our $H_0$ states that cases of a certain case type are assigned to the judges at random with the same baseline rate as the overall prevalence of that case type (and similar for entity labels). 
We calculate those base rates by averaging across all datasets and thus we introduce a dependency between the tests (which becomes, however, asymptotically small in the limit of large number of tests). 
The BY method is thus our preferred correction. 

To test predictability of a judge's cases (see also \ref{SI:judge_predictability_sensitivity}), our $H_0$ states that we cannot predict a judge's cases better than what would be expected from random guessing. 
The datasets are independent across the judges and thus BH is our preferred correction.

\subsection{Testing Random Assignment of Case Types}

To evaluate the randomness of case assignments, we compare the number of cases of each case type assigned to a given judge against the distribution of case types within that judge’s circuit during the relevant decade (see Methods). 
We utilize a binomial p-test to assess the null hypothesis that cases are randomly sampled from a binomial distribution.
If the assignment is indeed random, we expect the fraction of judges below any threshold p-value, $p$, to be less than or equal to $p$.

\begin{figure*}[!htb]
    \centering
    \includegraphics[width=\linewidth]{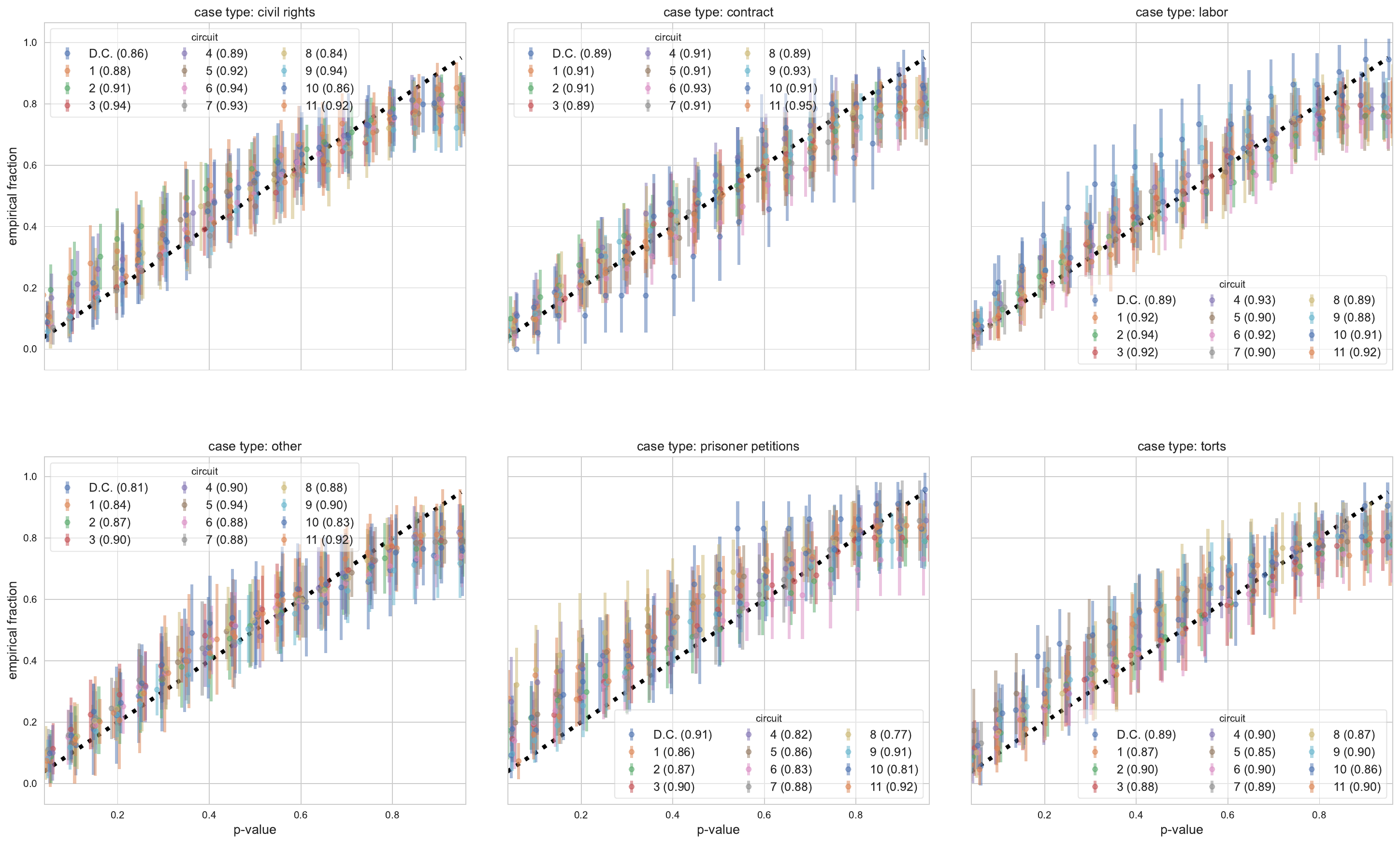}
    \caption{\textbf{Confirming random assignment for each case type with QQ plots.} In parenthesis we show the Pearson correlation coefficient between theoretical and empirical p-values.}
    \label{fig:case_type_bias_examination}
\end{figure*}

In Figure \ref{fig:case_type_bias_examination}, we plot this relationship for each case type and find strong evidence suggesting that the assignment is random, at least with respect to case types. 
Error bars in the plot are derived through bootstrapping the judges 100 times.

The p-values presented in Figure \ref{fig:case_type_bias_examination} were obtained across $K = 18,355$ distinct datasets, categorized by judge, epoch, and case type. 
As discussed earlier, these p-values necessitate adjustment for multiple hypothesis testing. 
After applying the Benjamini-Yekutieli (BY) correction, only $0.6\%$ of the corresponding corrected p-values fall below the $5\%$ threshold. Similarly, $1.6\%$ and $0.4\%$ of the p-values lie below the $5\%$ threshold when the Benjamini-Hochberg (BH) and Bonferroni (BF) corrections are applied, respectively.
These relatively low percentages indicate that, for the vast majority of jurisdictions, the null hypothesis of random case assignment cannot be rejected.

\subsection{Testing Random Assignment of Case Entities}

In the preceding analysis, we demonstrated that the hypothesis of random case assignment could not be rejected. 
Our evaluation primarily focuses on the types of cases assigned to judges. 
Ultimately, any analysis of random assignment will be based on observable attributes about cases.
To provide an additional check, we extend the analysis of random assignment beyond case types by using information about the identity of litigation parties from the CAP.

We categorized the cases into four distinct groups based on the nature of the parties involved:
(i) cases involving the government as one of the parties,
(ii) cases where both parties are companies,
(iii) cases where both parties are private individuals, and
(iv) cases with a mix of party types.
To systematically extract this information across all 112,312 cases, 
we employed OpenAI’s GPT-4 model, prompting it to assign each case to one of the four categories, or to return ``unknown'' if uncertain. 
The resulting distribution of labels is as follows: 
government cases (30\%), 
company cases (17\%), 
individual cases (10\%), 
mixed cases (36\%), 
and cases categorized as ``unknown'' (7\%). Manual inspection of 100 random records confirmed a $<5\%$ error rate for these labels.

\begin{figure*}[!htb]
    \centering
    \includegraphics[width=\linewidth]{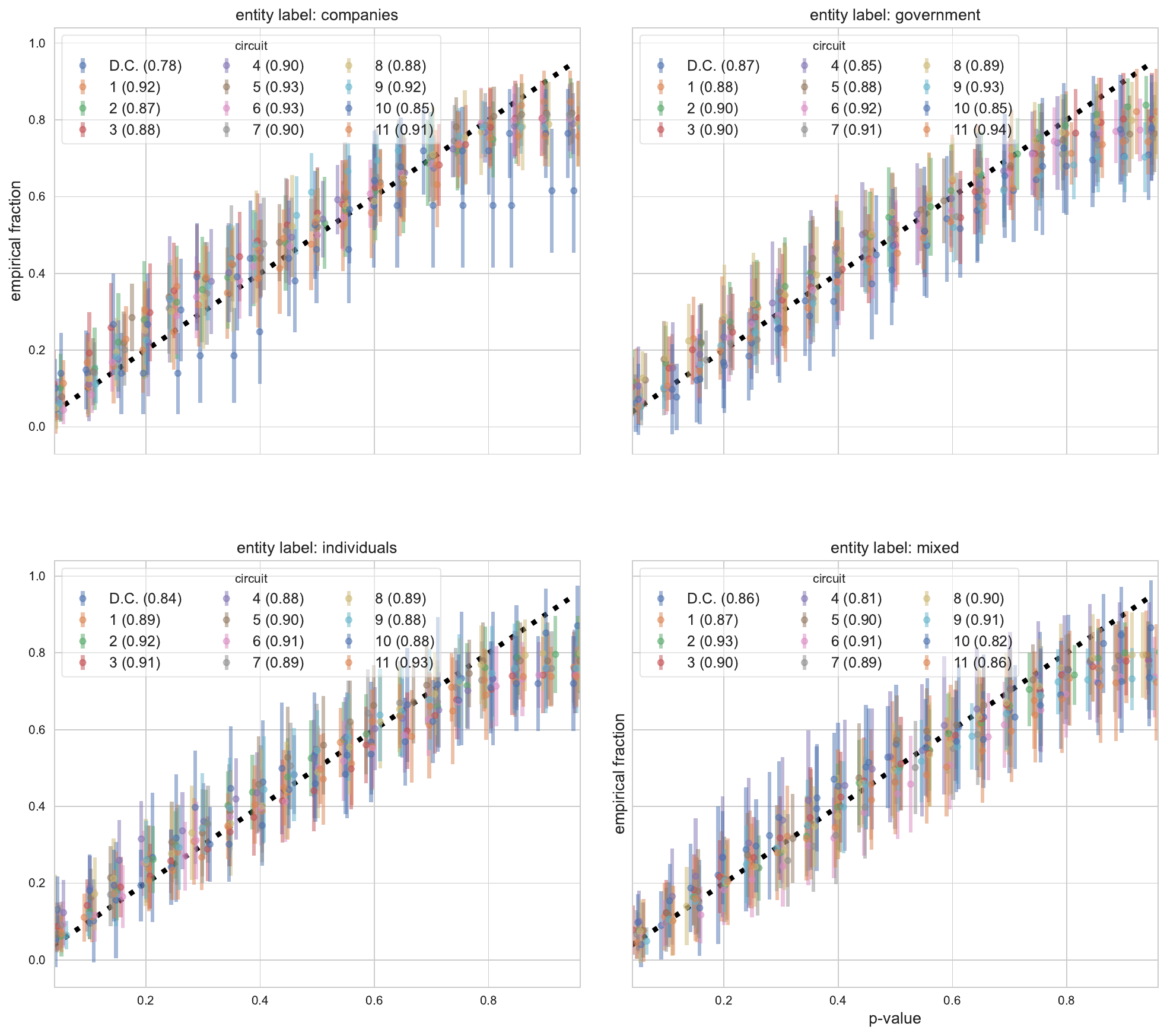}
    \caption{\textbf{Confirming random assignment for each litigant identity type with QQ plots.} In parenthesis we show the Pearson correlation coefficient between theoretical and empirical p-values.}
    \label{fig:entity_label_bias_examination}
\end{figure*}

We excluded the cases categorized as ``unknown'' and proceeded with the remaining cases, applying the same methodology used in our initial case-type analysis. 
Specifically, we grouped all cases by jurisdiction, decade, and judge, resulting in 15,133 distinct datasets. 
For each dataset, we tested the null hypothesis of random case assignment, comparing each group to the baseline average across all cases. After applying the Benjamini-Yekutieli (BY) correction, only 0.4\% of hypotheses were rejected 
(0.3\% with Bonferroni correction and 0.8\% with Benjamini-Hochberg correction), 
further supporting the inability to reject the null hypothesis of random case assignment. 
The corresponding QQ plots are presented in Figure \ref{fig:entity_label_bias_examination}.

\clearpage
\newpage
\section{Biographic Judge Features}
\label{SI:bio_features}

\begin{figure*}[!htb]
    \centering
    \includegraphics[width=0.8\linewidth]{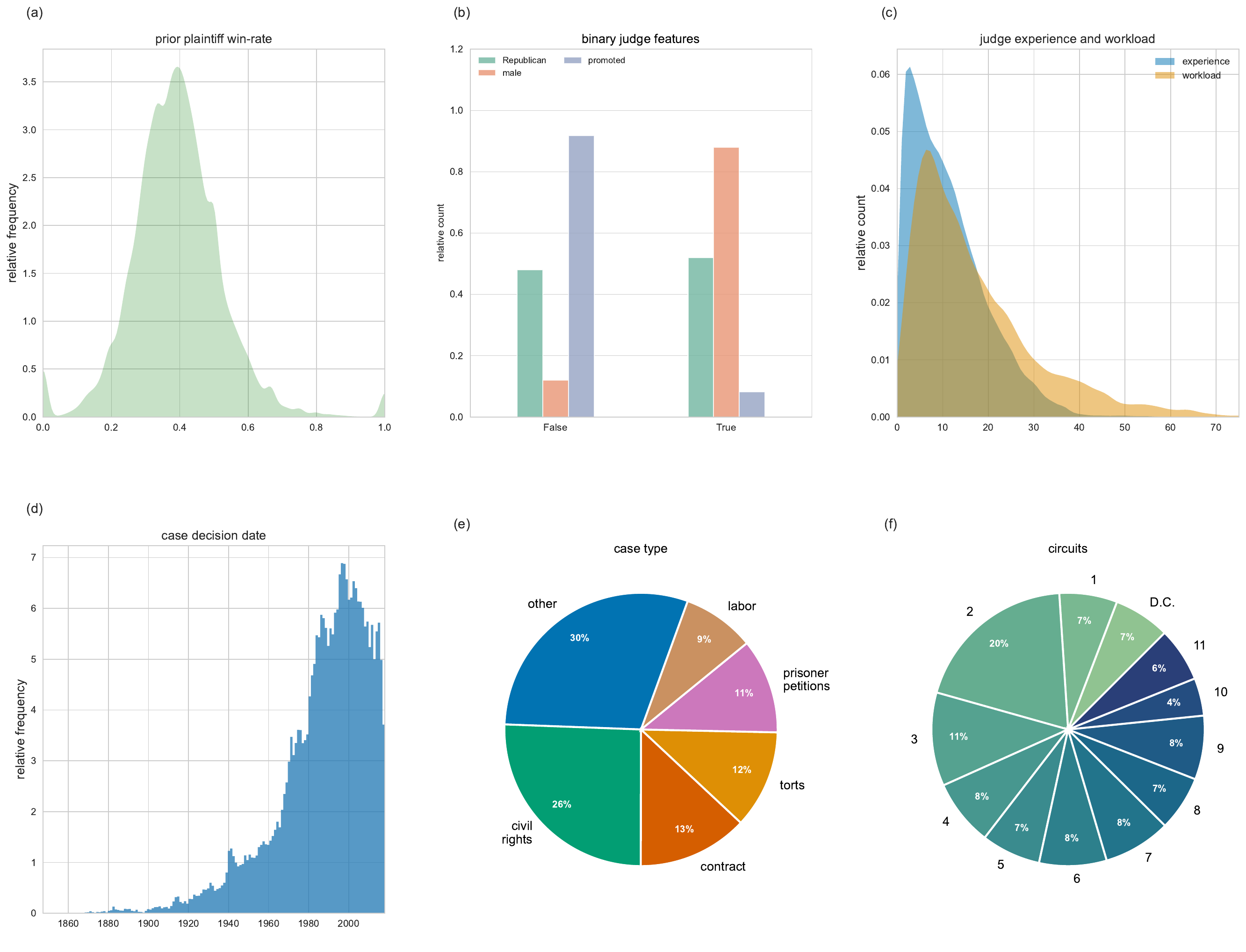}
    \caption{
            \textbf{Distribution of judge characteristics.}          
            (a)
            Plaintiff win-rate. 
            (b)
            Relative fraction of binary judge characteristics.            
            (c)
            Distribution of judge experience (in units of years since appointment) and judge workload (in units of average number of cases per year).
            (d)
            Distribution of case decision date. 
            (e) 
            Frequency of different case types.
            (f) 
            Frequency of different judicial circuits in which cases appear (these correspond to geography). 
            }
    \label{fig:feature_overview}
\end{figure*}

Figure \ref{fig:feature_overview} shows summary statistics of the biographic judge features.
Unless they are fixed effects, all features are re-calculated each time a judge is assigned a new case, relying on information that has been available 60 days prior to the case. 

\clearpage
\newpage
\section{Case Outcome Predictions: Sensitivity Analysis}
\label{SI:sensitivity_analysis}

Here we examine the sensitivity of our results with respect to different machine learning methods and data quality. 

\subsection{Machine Learning Models}

At the heart of this work lies the prediction of case outcomes as a function of judge characteristics.
For the prediction task, we have trained four machine learning models:
A gradient boost model (GB),
a random forest (RF), 
a multi-layer perceptron (MP), 
and a logistic regression (LR). 
For each model we use cross-validation to determine hyperparameters. 
For GB, hyperparameters include 
the number of estimators (25, 50, 100), 
maximum depth (2, 4, 5), 
and learning rate (0.01, 0.1, 0.2).
The RF model was optimized with 500 estimators. 
Hyperparameters include
maximum tree depth (2,3,4), 
minimum samples per leaf (0.01, 0.05, 0.10).
Hyperparameters for the MP model include 
with different hidden layer configurations, 
including $(n, n/2)$ and $(2n, n, n/2)$ where $n$ denotes the number of input features, 
along with varying L2 regularization values (0.0001, 0.001, 0.01). 
Finally, the LR model was evaluated with different L2 regularization strengths $(0.01, 0.1, 1, 10)$. 
To find the best hyperparameters, a 3-fold grid search was employed with accuracy as the evaluation metric.

\subsection{Data Quality}

Recall the ``classification confidence threshold'' $\tau$ from \ref{SI:transformer_labeling}. 
We note that increasing $\tau$ can be interpreted as increasing the quality of our dataset in the sense of having fewer mislabeled cases (see Figure \ref{fig:transformer_classifier_sensitivity}). 
Therefore, examining how our predictions with judge features depend on $\tau$ allows us to analyze the dependency of our results on data quality. 

While in \ref{SI:transformer_labeling} we test the accuracy of the transformer with the judge's opinion as input, 
here we test the accuracy of our four machine learning models with judges' biographic features as inputs. 
As such, the later case is much more difficult task than the former, since, in principle, all the information about the case outcome is contained in the judge's case summary. 

For fixed $\tau$, we consider all cases that the transformer labels with $p < 0.5 - \tau$ or $p > 0.5 + \tau$. 
We then down-sample the majority class (plaintiff lost the case) such that both the training set and the test set have an equal number of cases where the plaintiff won, resulting in a 50\% balance. 
This facilitates the interpretation and comparability of the prediction accuracy on the test set.

\begin{figure*}[!htb]
    \centering
    \includegraphics[width=0.8\linewidth]{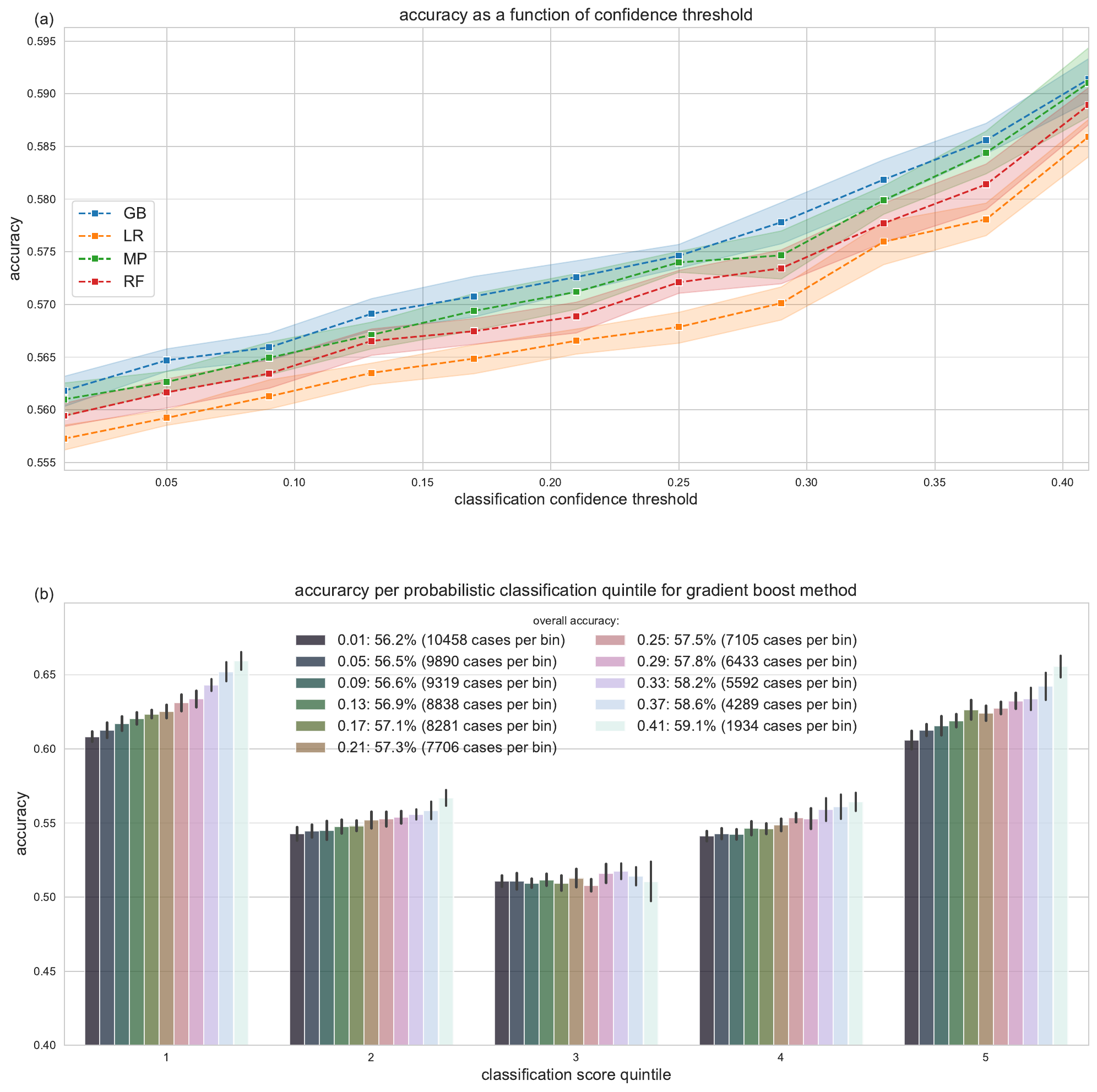}
    \caption{
            \textbf{Impact of different model choices on classification accuracy.}
            (top)
            Case outcome prediction accuracy as a function of the transformer's classification confidence threshold $\tau$ (as in Figure \ref{fig:transformer_classifier_sensitivity}) for four different machine learning methods: 
            Gradient boost (GB), 
            linear regression (LR), 
            multilayer perceptron (MP)
            and random forest (RF). 
            (bottom)
            Accuracy of the gradient boost probabilistic classifier segmented by classification confidence threshold. 
            Different colors indicate different confidence thresholds $\tau$.
            }
    \label{fig:case_prediction_sensitivity}
\end{figure*}

Next, we use the biographic judge features to train and test the machine learning algorithms on an 80\%-20\% train-test split. 
Figure \ref{fig:case_prediction_sensitivity} (top) illustrates the test accuracy for different values of $\tau$ across all four machine learning methods.
First, we observe that even for low values of $\tau$, the 50\% benchmark, which corresponds to random guessing, is significantly exceeded. Second, this tendency increases linearly with $\tau$, further confirming the robustness of our results: 
the better the quality of our dataset, the better we are at predicting case outcomes. 
As mentioned in \ref{SI:transformer_labeling}, we have set $\tau = 0.4$ for all results presented in the main paper.
Third, we observe that performance is consistent across methods, with logistic regression generally performing the worst and the gradient boosting model typically performing the best. 
Consequently, we have relied on the gradient boosting method to present the results in the main paper.

Finally, and akin to Figure 2 in the main paper, Figure \ref{fig:case_prediction_sensitivity} (bottom) shows the dependency of the classification results on the gradient boost model's confidence. Different colors represent different values of $\tau$. 
The monotonicity in $\tau$ is consistent with the top Figure, and serves as additional consistency check.

\clearpage
\newpage
\section{Logistic Regressions with Biographic Features}
\label{SI:bio_classification}

Here we classify case outcomes based on biographical features using logistic regressions as opposed to the gradient boost method presented in the main paper. 
We do not split the data into train- and test set.
Instead, we fit and study the regression coefficients across the entire dataset (Tables \ref{tab:civil_rights_logistic_bio}-\ref{tab:other_logistic_bio}).
We observe that the historical win rates are significant in each of these regressions, suggesting that judicial idiosyncrasies are persistent. 
The significance of judge's party affiliation varies between case types; republican judges display significantly lower win rates for torts, prisoner petitions, labor, and civil rights cases. 
We find that the effect of the circuit varies strongly between case types, that is, there does not appear to be a universally plaintiff-friendly or unfriendly circuit. 
Note that the ninth circuit is used as a reference value in the following tables, so as to avoid multi-collinearity. 

\begin{table}[!ht]
  \centering
  \begin{minipage}{0.49\textwidth}
    \centering
    \scriptsize
    \begin{tabular}{lcccccc}
    \toprule
    \textbf{judgement=1} & \textbf{coef} & \textbf{std err} & \textbf{z} & \textbf{P$>|$z$|$} & \textbf{[0.025} & \textbf{0.975]} \\
    \midrule
    \textbf{Intercept} & 0.3390 & 0.114 & 2.981 & 0.003 & 0.116 & 0.562 \\
    \textbf{decision\_date} & -0.3236 & 0.028 & -11.740 & 0.000 & -0.378 & -0.270 \\
    \textbf{experience} & -0.0063 & 0.024 & -0.259 & 0.796 & -0.054 & 0.041 \\
    \textbf{win\_rate} & 0.3428 & 0.025 & 13.515 & 0.000 & 0.293 & 0.393 \\
    \textbf{workload} & -0.1110 & 0.026 & -4.322 & 0.000 & -0.161 & -0.061 \\
    \textbf{D.C. circuit} & -0.3767 & 0.132 & -2.857 & 0.004 & -0.635 & -0.118 \\
    \textbf{circuit\_1} & -0.1479 & 0.130 & -1.138 & 0.255 & -0.402 & 0.107 \\
    \textbf{circuit\_2} & -0.5060 & 0.112 & -4.509 & 0.000 & -0.726 & -0.286 \\
    \textbf{circuit\_3} & -0.5606 & 0.125 & -4.484 & 0.000 & -0.806 & -0.316 \\
    \textbf{circuit\_4} & -0.6014 & 0.128 & -4.707 & 0.000 & -0.852 & -0.351 \\
    \textbf{circuit\_5} & -0.4964 & 0.128 & -3.863 & 0.000 & -0.748 & -0.245 \\
    \textbf{circuit\_6} & -0.3454 & 0.124 & -2.789 & 0.005 & -0.588 & -0.103 \\
    \textbf{circuit\_7} & -0.4593 & 0.124 & -3.717 & 0.000 & -0.702 & -0.217 \\
    \textbf{circuit\_8} & -0.2185 & 0.131 & -1.669 & 0.095 & -0.475 & 0.038 \\
    \textbf{circuit\_10} & -0.5507 & 0.142 & -3.876 & 0.000 & -0.829 & -0.272 \\
    \textbf{circuit\_11} & -0.4257 & 0.124 & -3.446 & 0.001 & -0.668 & -0.184 \\
    \textbf{gender\_male} & 0.1046 & 0.071 & 1.463 & 0.143 & -0.035 & 0.245 \\
    \textbf{party\_republican} & -0.2375 & 0.048 & -4.992 & 0.000 & -0.331 & -0.144 \\
    \textbf{promoted} & 0.0851 & 0.096 & 0.887 & 0.375 & -0.103 & 0.273 \\
    \midrule
    \textbf{Pseudo-R\textsuperscript{2}} & \textbf{6.6\%} & & & & & \\
    \bottomrule
    \end{tabular}
    \caption{\textbf{Logistic regression for civil rights cases.}}
    \label{tab:civil_rights_logistic_bio}
  \end{minipage}
  \hfill
  \begin{minipage}{0.49\textwidth}
    \centering
    \scriptsize
    \begin{tabular}{lcccccc}
    \toprule
    \textbf{judgement=1} & \textbf{coef} & \textbf{std err} & \textbf{z} & \textbf{P$>|$z$|$} & \textbf{[0.025} & \textbf{0.975]} \\
    \midrule
    \textbf{Intercept}         & -0.0665 & 0.097 & -0.688 & 0.491 & -0.256 & 0.123 \\
    \textbf{decision\_date}    & -0.1110 & 0.022 & -4.979 & 0.000 & -0.155 & -0.067 \\
    \textbf{experience}        & -0.0220 & 0.020 & -1.103 & 0.270 & -0.061 & 0.017 \\
    \textbf{win\_rate}         &  0.1759 & 0.021 &  8.437 & 0.000 &  0.135 & 0.217 \\
    \textbf{workload}          &  0.0115 & 0.021 &  0.540 & 0.589 & -0.030 & 0.053 \\
    \textbf{D.C. circuit}        & -0.1333 & 0.136 & -0.982 & 0.326 & -0.399 & 0.133 \\
    \textbf{circuit\_1}        & -0.0556 & 0.105 & -0.528 & 0.597 & -0.262 & 0.151 \\
    \textbf{circuit\_2}        &  0.0208 & 0.087 &  0.238 & 0.812 & -0.151 & 0.192 \\
    \textbf{circuit\_3}        &  0.1670 & 0.092 &  1.816 & 0.069 & -0.013 & 0.347 \\
    \textbf{circuit\_4}        &  0.1151 & 0.101 &  1.139 & 0.255 & -0.083 & 0.313 \\
    \textbf{circuit\_5}        &  0.0559 & 0.098 &  0.571 & 0.568 & -0.136 & 0.248 \\
    \textbf{circuit\_6}        &  0.1338 & 0.103 &  1.295 & 0.195 & -0.069 & 0.336 \\
    \textbf{circuit\_7}        & -0.2121 & 0.108 & -1.964 & 0.050 & -0.424 & 0.000 \\
    \textbf{circuit\_8}        &  0.2908 & 0.101 &  2.868 & 0.004 &  0.092 & 0.490 \\
    \textbf{circuit\_10}       &  0.2162 & 0.115 &  1.884 & 0.060 & -0.009 & 0.441 \\
    \textbf{circuit\_11}       &  0.1642 & 0.105 &  1.566 & 0.117 & -0.041 & 0.370 \\
    \textbf{gender\_male}      &  0.0448 & 0.069 &  0.646 & 0.518 & -0.091 & 0.181 \\
    \textbf{party\_republican} & -0.0875 & 0.040 & -2.211 & 0.027 & -0.165 & -0.010 \\
    \textbf{promoted}          & -0.0389 & 0.067 & -0.582 & 0.561 & -0.170 & 0.092 \\
    \midrule
    \textbf{Pseudo-R\textsuperscript{2}} & \textbf{1.4\%} & & & & & \\
    \bottomrule
    \end{tabular}
    \caption{\textbf{Logistic regression for contract cases.}}
    \label{tab:contract_logistic_bio}
  \end{minipage}
\end{table}

\begin{table}[!ht]
  \centering
  \begin{minipage}{0.49\textwidth}
    \centering
    \scriptsize
    \begin{tabular}{lcccccc}
    \toprule
    \textbf{judgement=1} & \textbf{coef} & \textbf{std err} & \textbf{z} & \textbf{P$>|$z$|$} & \textbf{[0.025} & \textbf{0.975]} \\
    \midrule
    \textbf{Intercept}         & 0.0448 & 0.177 & 0.253 & 0.800 & -0.302 & 0.392 \\
    \textbf{decision\_date}    & 0.1600 & 0.048 & 3.335 & 0.001 & 0.066 & 0.254 \\
    \textbf{experience}        & -0.0827 & 0.044 & -1.879 & 0.060 & -0.169 & 0.004 \\
    \textbf{win\_rate}         & 0.4000 & 0.045 & 8.892 & 0.000 & 0.312 & 0.488 \\
    \textbf{workload}          & -0.3679 & 0.051 & -7.217 & 0.000 & -0.468 & -0.268 \\
    \textbf{D.C. circuit}        & -0.6635 & 0.283 & -2.343 & 0.019 & -1.219 & -0.108 \\
    \textbf{circuit\_1}        & -0.2554 & 0.212 & -1.206 & 0.228 & -0.671 & 0.160 \\
    \textbf{circuit\_2}        & -0.5539 & 0.163 & -3.401 & 0.001 & -0.873 & -0.235 \\
    \textbf{circuit\_3}        & -0.5867 & 0.194 & -3.029 & 0.002 & -0.966 & -0.207 \\
    \textbf{circuit\_4}        & -0.7771 & 0.184 & -4.229 & 0.000 & -1.137 & -0.417 \\
    \textbf{circuit\_5}        & 0.1516 & 0.221 & 0.687 & 0.492 & -0.281 & 0.584 \\
    \textbf{circuit\_6}        & -0.3800 & 0.189 & -2.010 & 0.044 & -0.750 & -0.009 \\
    \textbf{circuit\_7}        & -0.1593 & 0.198 & -0.806 & 0.420 & -0.547 & 0.228 \\
    \textbf{circuit\_8}        & -0.4529 & 0.192 & -2.353 & 0.019 & -0.830 & -0.076 \\
    \textbf{circuit\_10}       & -0.5853 & 0.223 & -2.626 & 0.009 & -1.022 & -0.149 \\
    \textbf{circuit\_11}       & 0.3505 & 0.244 & 1.437 & 0.151 & -0.128 & 0.828 \\
    \textbf{gender\_male}      & 0.3719 & 0.141 & 2.632 & 0.008 & 0.095 & 0.649 \\
    \textbf{party\_republican} & -0.2700 & 0.088 & -3.073 & 0.002 & -0.442 & -0.098 \\
    \textbf{promoted}          & -0.1260 & 0.163 & -0.775 & 0.438 & -0.445 & 0.193 \\
    \midrule
    \textbf{Pseudo-R\textsuperscript{2}} & \textbf{8.7\%} & & & & & \\
    \bottomrule
    \end{tabular}
    \caption{\textbf{Logistic regression for prisoner petition cases.}}
    \label{tab:prisoner_logistic_bio}
  \end{minipage}
  \hfill
  \begin{minipage}{0.49\textwidth}
    \centering
    \scriptsize
    \begin{tabular}{lcccccc}
    \toprule
    \textbf{judgement=1} & \textbf{coef} & \textbf{std err} & \textbf{z} & \textbf{P$>|$z$|$} & \textbf{[0.025} & \textbf{0.975]} \\
    \midrule
    \textbf{Intercept}         & -0.4831 & 0.161 & -3.006 & 0.003 & -0.798 & -0.168 \\
    \textbf{decision\_date}    & -0.2810 & 0.032 & -8.831 & 0.000 & -0.343 & -0.219 \\
    \textbf{experience}        & -0.0268 & 0.028 & -0.965 & 0.334 & -0.081 & 0.028 \\
    \textbf{win\_rate}         & 0.2718 & 0.030 & 9.172 & 0.000 & 0.214 & 0.330 \\
    \textbf{workload}          & -0.0027 & 0.030 & -0.089 & 0.929 & -0.062 & 0.057 \\
    \textbf{D.C. circuit}        & 0.8469 & 0.174 & 4.862 & 0.000 & 0.505 & 1.188 \\
    \textbf{circuit\_1}        & 0.0173 & 0.161 & 0.108 & 0.914 & -0.297 & 0.332 \\
    \textbf{circuit\_2}        & 0.0433 & 0.135 & 0.320 & 0.749 & -0.222 & 0.308 \\
    \textbf{circuit\_3}        & 0.2590 & 0.134 & 1.926 & 0.054 & -0.005 & 0.522 \\
    \textbf{circuit\_4}        & -0.1341 & 0.153 & -0.879 & 0.380 & -0.433 & 0.165 \\
    \textbf{circuit\_5}        & 0.3561 & 0.138 & 2.590 & 0.010 & 0.087 & 0.626 \\
    \textbf{circuit\_6}        & -0.1653 & 0.155 & -1.063 & 0.288 & -0.470 & 0.139 \\
    \textbf{circuit\_7}        & -0.2186 & 0.178 & -1.226 & 0.220 & -0.568 & 0.131 \\
    \textbf{circuit\_8}        & 0.2961 & 0.161 & 1.839 & 0.066 & -0.020 & 0.612 \\
    \textbf{circuit\_10}       & -0.1346 & 0.192 & -0.702 & 0.483 & -0.511 & 0.241 \\
    \textbf{circuit\_11}       & -0.1354 & 0.166 & -0.817 & 0.414 & -0.460 & 0.190 \\
    \textbf{gender\_male}      & 0.5162 & 0.119 & 4.346 & 0.000 & 0.283 & 0.749 \\
    \textbf{party\_republican} & -0.3279 & 0.056 & -5.839 & 0.000 & -0.438 & -0.218 \\
    \textbf{promoted}          & 0.0295 & 0.095 & 0.312 & 0.755 & -0.156 & 0.215 \\
    \midrule
    \textbf{Pseudo-R\textsuperscript{2}} & \textbf{5.5\%} & & & & & \\
    \bottomrule
    \end{tabular}
    \caption{\textbf{Logistic regression for tort cases.}}
    \label{tab:tort_logistic_bio}
  \end{minipage}
\end{table}

\begin{table}
  \centering
  \begin{minipage}{0.49\textwidth}
    \centering
    \scriptsize
    \begin{tabular}{lcccccc}
    \toprule
    \textbf{judgement=1} & \textbf{coef} & \textbf{std err} & \textbf{z} & \textbf{P$>|$z$|$} & \textbf{[0.025} & \textbf{0.975]} \\
    \midrule
    \textbf{Intercept}         & 0.0568 & 0.123 & 0.462 & 0.644 & -0.184 & 0.298 \\
    \textbf{decision\_date}    & -0.0260 & 0.031 & -0.845 & 0.398 & -0.086 & 0.034 \\
    \textbf{experience}        & 0.0070 & 0.028 & 0.255 & 0.799 & -0.047 & 0.061 \\
    \textbf{win\_rate}         & 0.1730 & 0.029 & 6.054 & 0.000 & 0.117 & 0.229 \\
    \textbf{workload}          & -0.0438 & 0.030 & -1.458 & 0.145 & -0.103 & 0.015 \\
    \textbf{D.C. circuit}        & 0.4184 & 0.153 & 2.734 & 0.006 & 0.118 & 0.718 \\
    \textbf{circuit\_1}        & -0.4425 & 0.146 & -3.021 & 0.003 & -0.730 & -0.155 \\
    \textbf{circuit\_2}        & -0.0516 & 0.119 & -0.432 & 0.666 & -0.286 & 0.183 \\
    \textbf{circuit\_3}        & -0.1572 & 0.126 & -1.248 & 0.212 & -0.404 & 0.090 \\
    \textbf{circuit\_4}        & -0.1860 & 0.135 & -1.373 & 0.170 & -0.451 & 0.079 \\
    \textbf{circuit\_5}        & -0.1273 & 0.144 & -0.886 & 0.376 & -0.409 & 0.154 \\
    \textbf{circuit\_6}        & -0.2006 & 0.126 & -1.593 & 0.111 & -0.447 & 0.046 \\
    \textbf{circuit\_7}        & -0.0463 & 0.135 & -0.345 & 0.730 & -0.310 & 0.217 \\
    \textbf{circuit\_8}        & -0.0668 & 0.137 & -0.489 & 0.624 & -0.334 & 0.201 \\
    \textbf{circuit\_10}       & -0.2335 & 0.177 & -1.318 & 0.187 & -0.581 & 0.114 \\
    \textbf{circuit\_11}       & -0.0746 & 0.141 & -0.530 & 0.596 & -0.350 & 0.201 \\
    \textbf{gender\_male}      & 0.0993 & 0.087 & 1.136 & 0.256 & -0.072 & 0.271 \\
    \textbf{party\_republican} & -0.1103 & 0.055 & -1.998 & 0.046 & -0.218 & -0.002 \\
    \textbf{promoted}          & 0.0288 & 0.104 & 0.277 & 0.781 & -0.175 & 0.232 \\
    \midrule
    \textbf{Pseudo-R\textsuperscript{2}} & \textbf{1.2\%} & & & & & \\
    \bottomrule
    \end{tabular}
    \caption{\textbf{Logistic regression for labor cases.}}
    \label{tab:labor_logistic_bio}
  \end{minipage}
  \hfill
  \begin{minipage}{0.49\textwidth}
    \centering
    \scriptsize
    \begin{tabular}{lcccccc}
    \toprule
    \textbf{judgement=1} & \textbf{coef} & \textbf{std err} & \textbf{z} & \textbf{P$>|$z$|$} & \textbf{[0.025} & \textbf{0.975]} \\
    \midrule
    \textbf{Intercept}         & -0.2101 & 0.064 & -3.270 & 0.001 & -0.336 & -0.084 \\
    \textbf{decision\_date}    & -0.0903 & 0.017 & -5.463 & 0.000 & -0.123 & -0.058 \\
    \textbf{experience}        &  0.0103 & 0.015 &  0.707 & 0.480 & -0.018 &  0.039 \\
    \textbf{win\_rate}         &  0.1889 & 0.016 & 12.156 & 0.000 &  0.158 &  0.219 \\
    \textbf{workload}          & -0.0610 & 0.016 & -3.800 & 0.000 & -0.092 & -0.030 \\
    \textbf{D.C. circuit}        & -0.3582 & 0.071 & -5.065 & 0.000 & -0.497 & -0.220 \\
    \textbf{circuit\_1}        &  0.0884 & 0.073 &  1.207 & 0.227 & -0.055 &  0.232 \\
    \textbf{circuit\_2}        &  0.0488 & 0.058 &  0.840 & 0.401 & -0.065 &  0.163 \\
    \textbf{circuit\_3}        &  0.0022 & 0.064 &  0.035 & 0.972 & -0.124 &  0.128 \\
    \textbf{circuit\_4}        &  0.2527 & 0.074 &  3.430 & 0.001 &  0.108 &  0.397 \\
    \textbf{circuit\_5}        & -0.0204 & 0.080 & -0.255 & 0.798 & -0.177 &  0.136 \\
    \textbf{circuit\_6}        &  0.0074 & 0.072 &  0.103 & 0.918 & -0.133 &  0.148 \\
    \textbf{circuit\_7}        & -0.0255 & 0.072 & -0.354 & 0.723 & -0.166 &  0.115 \\
    \textbf{circuit\_8}        &  0.2168 & 0.075 &  2.875 & 0.004 &  0.069 &  0.365 \\
    \textbf{circuit\_10}       &  0.0932 & 0.084 &  1.114 & 0.265 & -0.071 &  0.257 \\
    \textbf{circuit\_11}       &  0.3502 & 0.076 &  4.584 & 0.000 &  0.200 &  0.500 \\
    \textbf{gender\_male}      &  0.2081 & 0.050 &  4.189 & 0.000 &  0.111 &  0.305 \\
    \textbf{party\_republican} & -0.0458 & 0.030 & -1.546 & 0.122 & -0.104 &  0.012 \\
    \textbf{promoted}          & -0.0454 & 0.056 & -0.818 & 0.413 & -0.154 &  0.063 \\
    \midrule
    \textbf{Pseudo-R\textsuperscript{2}} & \textbf{1.9\%} & & & & & \\
    \bottomrule
    \end{tabular}
    \caption{\textbf{Logistic regression for ``other'' cases.}}
    \label{tab:other_logistic_bio}
  \end{minipage}
\end{table}

\clearpage
\newpage
\section{Shapley Feature Importances}
\label{SI:bio_shap}

\begin{figure*}[!htb]
    \centering
    \includegraphics[width=\linewidth]{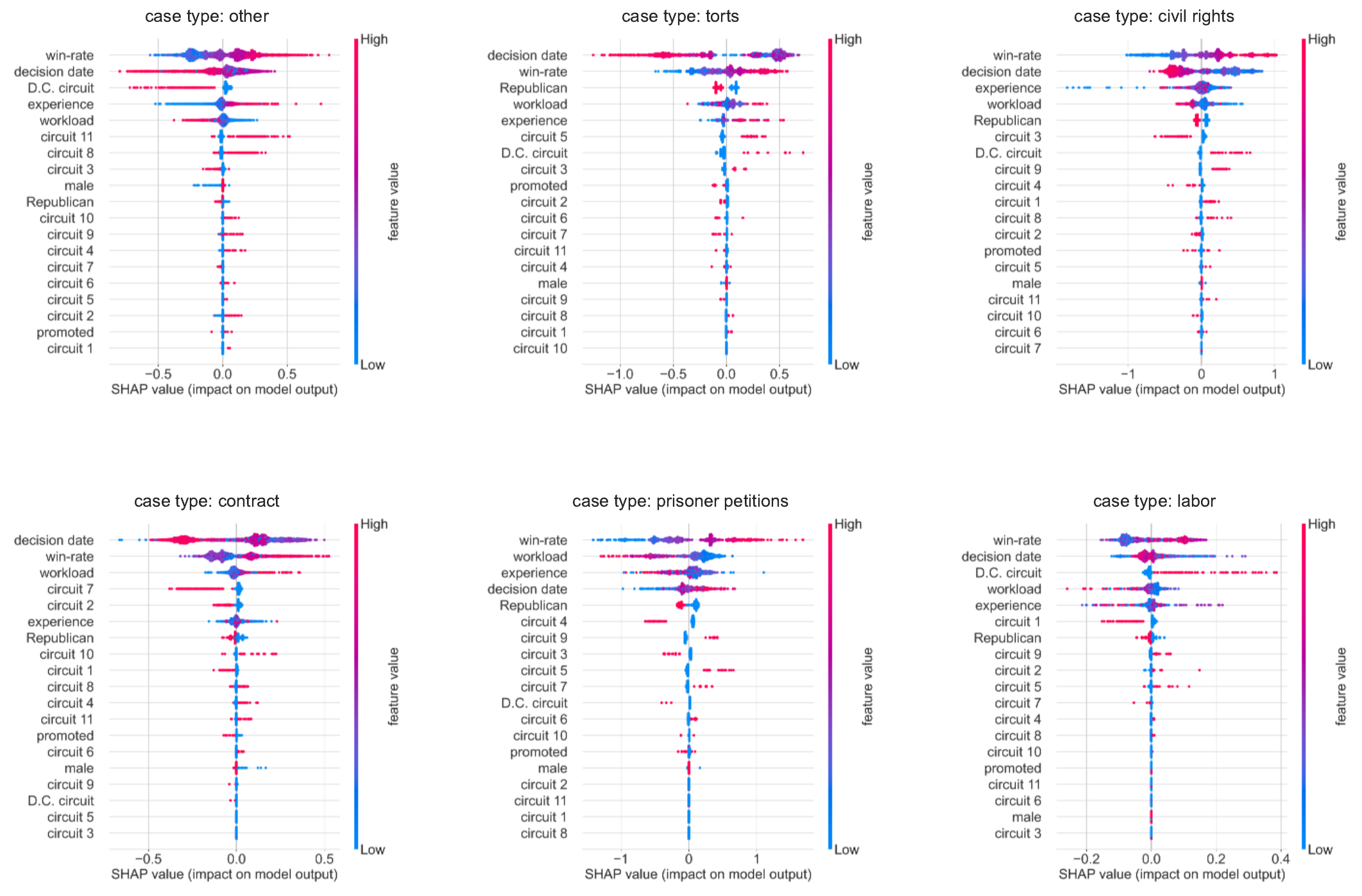}
    \caption{\textbf{Shapley feature importance for biographic classifier across all case types.}}
    \label{fig:bio_features_SHAP_overview}
\end{figure*}

To highlight the contribution of various features, we show Shapley feature importances.
Figure \ref{fig:bio_features_SHAP_overview} is an extension of Figure 3 in the main paper across different case types. 
We note that across all case types, judges' historical win rate is among the most important features, suggesting that judicial idiosyncrasies persist over judges' careers and motivating our investigation into early-career citation decisions.

\begin{figure*}[!htb]
    \centering
    \includegraphics[width=0.4\linewidth]{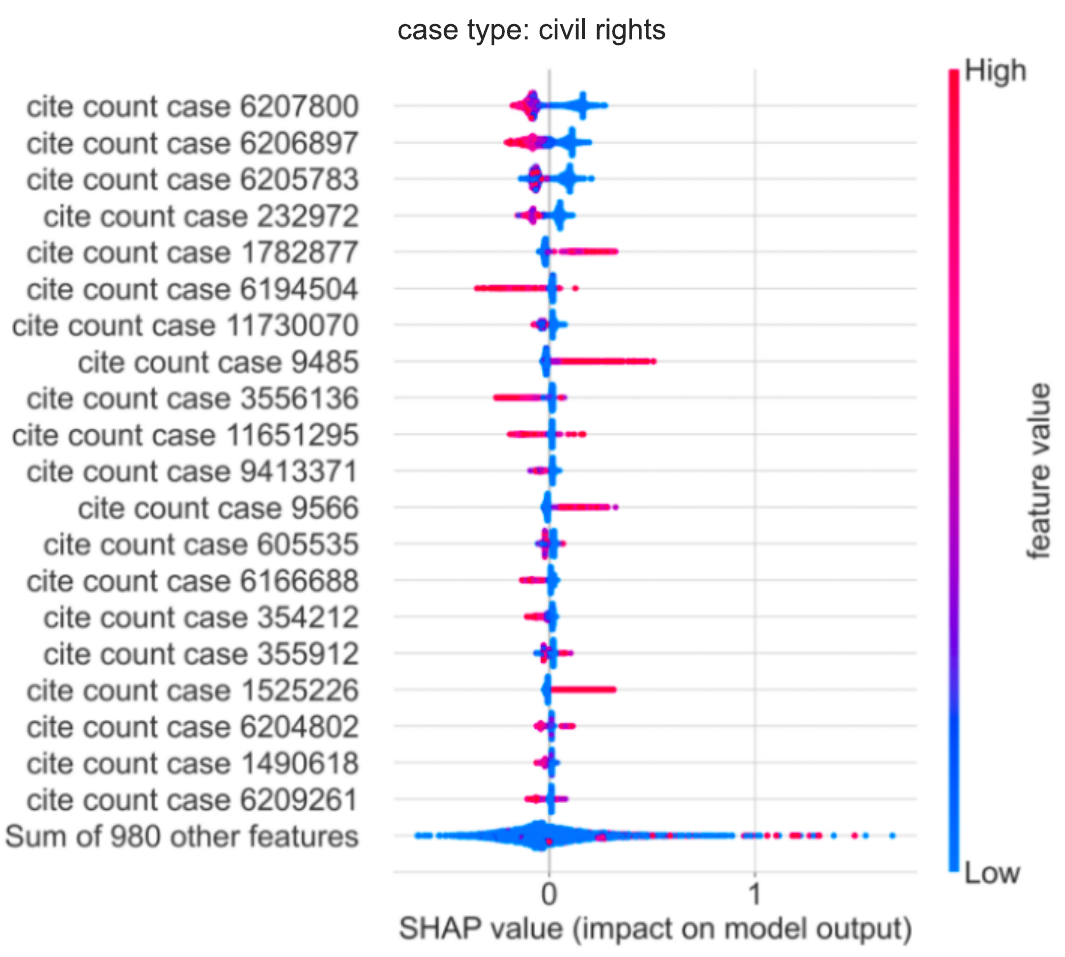}
    \caption{\textbf{Shapley feature importance for GB model trained on raw citation counts.} This plot illustrates the most important cases in the early-career citation records.}
    \label{fig:cite_features_civil_rights}
\end{figure*}

Figure \ref{fig:cite_features_civil_rights} shows the Shapley importances for a gradient boost (GB) model that has been trained on the raw citation features, as opposed to first reducing the dimensionality via NMF embedding. 
Two of the more prominent contributions are coming from citations to the cases with ID's 6207800 and 6206897 that have been explicitly discussed in the main text. 

\clearpage
\newpage
\section{Predicting Judge Characteristics using Citations}
\label{SI:predict_bio_with_cites}

Here we test the degree to which early-career citation features explain judge's biographic features.
To this end, we present a regression of the six biographic features against the 30-dimensional NMF citation features across all available cases.
To avoid multi-collinearity we employ a regularization value of $0.001$.
For continuous biographic features (experience, historical win rate, and workload) we use a linear regression and report $R^2$ values.
For the binary features (gender, party affiliation, and promotions) we use a logistic regression and report pseudo-$R^2$ values.
We find that citations capture a remarkable variance in biographic features (up to 23\% for the ``promoted'' feature). 
To test robustness, we confirm that replacing citation features by Gaussian random noise results in quasi zero (pseudo)-$R^2$ for all regressions.

\begin{table}[ht]
\centering
\begin{tabular}{lcccccc}
\toprule
& experience & gender male & party republican & promoted & win rate & workload \\
\midrule
\textbf{(Pseudo-) R\textsuperscript{2}} & 4.8\% & 23.1\% & 13.3\% & 23.3\% & 14.2\% & 7.6\% \\
\midrule\midrule
Intercept & 0.11*** (0.00) & 4.97*** (0.00) & -0.04** (0.00) & -2.93*** (0.00) & 0.06*** (0.00) & 0.03*** (0.00) \\
NMF dim 0 & -0.01 (0.39) & -0.10** (0.01) & 0.25*** (0.00) & -0.55*** (0.00) & -0.13*** (0.00) & -0.12*** (0.00) \\
NMF dim 1 & -0.04*** (0.00) & -0.06* (0.04) & -0.32*** (0.00) & -0.26*** (0.00) & -0.13*** (0.00) & 0.15*** (0.00) \\
NMF dim 2 & -0.01 (0.08) & 0.24*** (0.00) & -0.33*** (0.00) & -0.02 (0.26) & -0.14*** (0.00) & 0.15*** (0.00) \\
NMF dim 3 & -0.07*** (0.00) & 0.08* (0.04) & 0.38*** (0.00) & 0.06 (0.38) & -0.22*** (0.00) & 0.06*** (0.00) \\
NMF dim 4 & -0.05*** (0.00) & 1.96*** (0.00) & -0.44*** (0.00) & 0.02 (0.09) & -0.01* (0.04) & -0.04*** (0.00) \\
NMF dim 5 & -0.02*** (0.00) & 0.00 (0.91) & 0.02 (0.25) & -0.19*** (0.00) & -0.05*** (0.00) & -0.01* (0.04) \\
NMF dim 6 & -0.02* (0.05) & -0.10*** (0.00) & -0.21*** (0.00) & -0.31*** (0.00) & -0.03*** (0.00) & -0.05*** (0.00) \\
NMF dim 7 & -0.03*** (0.00) & 0.17* (0.01) & 0.01 (0.38) & -0.20*** (0.00) & -0.15*** (0.00) & 0.09*** (0.00) \\
NMF dim 8 & 0.02*** (0.00) & -0.05 (0.19) & -0.03* (0.01) & -0.13*** (0.00) & -0.02*** (0.00) & -0.02** (0.00) \\
NMF dim 9 & -0.01 (0.15) & 0.61** (0.00) & 0.11*** (0.00) & -0.18*** (0.00) & -0.05*** (0.00) & -0.01 (0.30) \\
NMF dim 10 & -0.15*** (0.00) & -0.05** (0.00) & -0.50*** (0.00) & -0.00 (0.86) & -0.08*** (0.00) & -0.02** (0.00) \\
NMF dim 11 & -0.00 (0.57) & 0.58 (0.96) & 0.06 (0.12) & -0.59*** (0.00) & -0.02*** (0.00) & -0.01 (0.07) \\
NMF dim 12 & -0.04*** (0.00) & 1.22*** (0.00) & -0.07*** (0.00) & -0.02 (0.25) & -0.04*** (0.00) & 0.00 (0.41) \\
NMF dim 13 & -0.05*** (0.00) & 0.26*** (0.00) & 0.14*** (0.00) & -0.20*** (0.00) & -0.09*** (0.00) & -0.13*** (0.00) \\
NMF dim 14 & -0.02** (0.00) & 0.08* (0.02) & -0.19*** (0.00) & -0.40*** (0.00) & -0.00 (0.73) & -0.07*** (0.00) \\
NMF dim 15 & -0.00 (0.20) & 3.01 (0.24) & -0.04*** (0.00) & -0.06* (0.02) & 0.01** (0.00) & -0.01*** (0.00) \\
NMF dim 16 & -0.06*** (0.00) & 0.10** (0.00) & -0.09*** (0.00) & -0.13*** (0.00) & -0.13*** (0.00) & 0.11*** (0.00) \\
NMF dim 17 & -0.09*** (0.00) & -0.23*** (0.00) & -0.46*** (0.00) & -0.40*** (0.00) & -0.20*** (0.00) & 0.11*** (0.00) \\
NMF dim 18 & -0.01 (0.34) & 0.05 (0.09) & -0.05*** (0.00) & -0.15*** (0.00) & -0.01* (0.04) & -0.04*** (0.00) \\
NMF dim 19 & -0.05*** (0.00) & 2.16*** (0.00) & 0.13*** (0.00) & -0.09*** (0.00) & -0.05*** (0.00) & -0.01** (0.01) \\
NMF dim 20 & -0.01 (0.30) & 0.02 (0.48) & 0.17*** (0.00) & -0.37*** (0.00) & -0.12*** (0.00) & 0.13*** (0.00) \\
NMF dim 21 & 0.00 (0.56) & 0.17 (0.06) & -0.22*** (0.00) & 0.01 (0.41) & -0.02*** (0.00) & -0.03*** (0.00) \\
NMF dim 22 & 0.05*** (0.00) & 0.02 (0.60) & 0.10*** (0.00) & -0.20*** (0.00) & -0.06*** (0.00) & 0.08*** (0.00) \\
NMF dim 23 & -0.00 (0.63) & 0.04 (0.10) & -0.26*** (0.00) & -0.06*** (0.00) & -0.01** (0.00) & -0.02*** (0.00) \\
NMF dim 24 & 0.03*** (0.00) & 0.16 (0.16) & 0.02 (0.09) & -0.22*** (0.00) & -0.02*** (0.00) & -0.03*** (0.00) \\
NMF dim 25 & -0.03*** (0.00) & 0.07 (0.16) & -0.18*** (0.00) & 0.02 (0.44) & -0.06*** (0.00) & -0.03*** (0.00) \\
NMF dim 26 & -0.04*** (0.00) & 5.08*** (0.00) & 0.06** (0.00) & -0.24*** (0.00) & -0.08*** (0.00) & 0.05*** (0.00) \\
NMF dim 27 & 0.03*** (0.00) & 0.10*** (0.00) & -0.09*** (0.00) & -0.08** (0.01) & -0.09*** (0.00) & 0.11*** (0.00) \\
NMF dim 28 & -0.05*** (0.00) & 0.23 (0.12) & -0.02 (0.08) & -0.03 (0.06) & -0.01** (0.01) & -0.01* (0.02) \\
NMF dim 29 & 0.01 (0.36) & 0.27*** (0.00) & 0.40*** (0.00) & -0.33*** (0.00) & -0.06*** (0.00) & 0.02** (0.00) \\
\bottomrule
\end{tabular}
\caption{\textbf{Regression of 6 different judge characteristics against 30-dimensional NMF embeddings of early-career citation histories.} Significant $p$-values are annotated with one, two or three starts corresponding to significance a the $0.05, 0.01, 0.001$ level, respectively. The values indicated in brackets correspond to the $p$-values.
}
\label{tab:judge_feats_vs_NMF_feat}
\end{table}

\clearpage
\newpage
\section{Relationship between Prediction Accuracy and NMF Embedding Size}
\label{SI:prediction_accuracy_versus_embedding_dimensions}

\begin{figure*}[!htb]
    \centering
    \includegraphics[width=\linewidth]{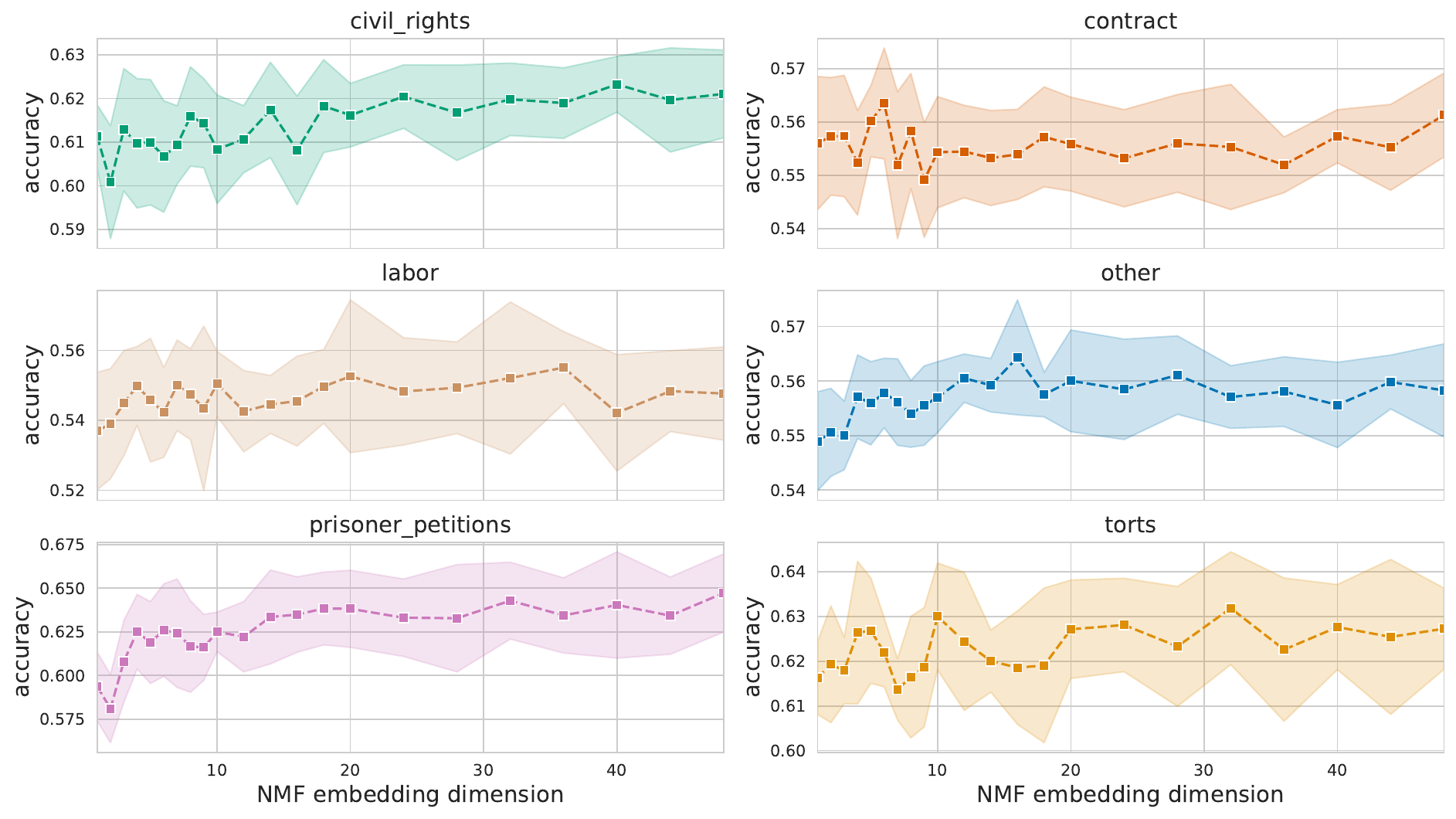}
    \caption{\textbf{Relationship between citation embedding size and classifier performance across case types.}}
    \label{fig:NMF_dim_dependency}
\end{figure*}

For each case type, we train gradient boost binary classifiers based on NMF embeddings of early citation counts. 
This is exactly as described in the Methods section. 
However, in Figure \ref{fig:NMF_dim_dependency} we vary the NMF embedding dimension from 1 to 50. 
We observe relatively fast convergence, suggesting that the citation features capture relatively low-dimensional, latent variables.
The results presented in the main paper are shown for an embedding dimension of 30. 

\clearpage
\newpage
\section{Logistic Regressions with Citation Features}
\label{SI:cite_classification}

Here we repeat the classification using embeddings of early-career citation histories and logistic regressions. 
We note that, while controlling for circuits and decision dates, some of the citation features are statistically significant suggesting that they explain future judicial decisions.
Note that the ninth circuit is used as a reference value in the following tables.

\begin{figure}[!ht]
  \centering
  \begin{minipage}{0.49\textwidth}
    \centering
    \scriptsize
    
    \begin{tabular}{lcccccc}
    \toprule
    \textbf{judgement=1} & \textbf{coef} & \textbf{std err} & \textbf{z} & \textbf{P$>|$z$|$} & \textbf{[0.025} & \textbf{0.975]} \\
    \midrule
    \textbf{Intercept}   & 0.4314  & 0.105  & 4.125  & 0.000  & 0.226  & 0.636  \\
    \textbf{NMF dim 0}   & -0.2418 & 0.057  & -4.267 & 0.000  & -0.353 & -0.131 \\
    \textbf{NMF dim 1}   & -0.1313 & 0.033  & -4.023 & 0.000  & -0.195 & -0.067 \\
    \textbf{NMF dim 2}   & -0.0800 & 0.057  & -1.406 & 0.160  & -0.191 & 0.032  \\
    \textbf{NMF dim 3}   & -0.0529 & 0.030  & -1.740 & 0.082  & -0.113 & 0.007  \\
    \textbf{NMF dim 4}   & -0.0897 & 0.037  & -2.455 & 0.014  & -0.161 & -0.018 \\
    \textbf{NMF dim 5}   & -0.0803 & 0.030  & -2.678 & 0.007  & -0.139 & -0.022 \\
    \textbf{NMF dim 6}   & 0.0029  & 0.030  & 0.095  & 0.924  & -0.056 & 0.062  \\
    \textbf{NMF dim 7}   & -0.0618 & 0.031  & -2.022 & 0.043  & -0.122 & -0.002 \\
    \textbf{NMF dim 8}   & -0.1405 & 0.041  & -3.388 & 0.001  & -0.222 & -0.059 \\
    \textbf{NMF dim 9}   & -0.1139 & 0.042  & -2.724 & 0.006  & -0.196 & -0.032 \\
    \textbf{NMF dim 10}  & -0.1555 & 0.030  & -5.147 & 0.000  & -0.215 & -0.096 \\
    \textbf{NMF dim 11}  & -0.0533 & 0.028  & -1.910 & 0.056  & -0.108 & 0.001  \\
    \textbf{NMF dim 12}  & -0.0877 & 0.035  & -2.514 & 0.012  & -0.156 & -0.019 \\
    \textbf{NMF dim 13}  & -0.0070 & 0.036  & -0.195 & 0.845  & -0.078 & 0.063  \\
    \textbf{NMF dim 14}  & 0.0174  & 0.028  & 0.613  & 0.540  & -0.038 & 0.073  \\
    \textbf{NMF dim 15}  & -0.0868 & 0.031  & -2.774 & 0.006  & -0.148 & -0.025 \\
    \textbf{NMF dim 16}  & -0.0205 & 0.029  & -0.712 & 0.476  & -0.077 & 0.036  \\
    \textbf{NMF dim 17}  & -0.0351 & 0.030  & -1.166 & 0.244  & -0.094 & 0.024  \\
    \textbf{NMF dim 18}  & -0.0784 & 0.030  & -2.653 & 0.008  & -0.136 & -0.020 \\
    \textbf{NMF dim 19}  & -0.0493 & 0.032  & -1.540 & 0.124  & -0.112 & 0.013  \\
    \textbf{NMF dim 20}  & -0.0484 & 0.030  & -1.636 & 0.102  & -0.106 & 0.010  \\
    \textbf{NMF dim 21}  & 0.0275  & 0.034  & 0.809  & 0.418  & -0.039 & 0.094  \\
    \textbf{NMF dim 22}  & 0.0547  & 0.028  & 1.956  & 0.050  & -0.000 & 0.109  \\
    \textbf{NMF dim 23}  & -0.1176 & 0.030  & -3.959 & 0.000  & -0.176 & -0.059 \\
    \textbf{NMF dim 24}  & -0.2043 & 0.047  & -4.305 & 0.000  & -0.297 & -0.111 \\
    \textbf{NMF dim 25}  & -0.0009 & 0.027  & -0.035 & 0.972  & -0.053 & 0.051  \\
    \textbf{NMF dim 26}  & -0.0243 & 0.027  & -0.887 & 0.375  & -0.078 & 0.029  \\
    \textbf{NMF dim 27}  & -0.0712 & 0.032  & -2.226 & 0.026  & -0.134 & -0.009 \\
    \textbf{NMF dim 28}  & -0.0854 & 0.032  & -2.649 & 0.008  & -0.149 & -0.022 \\
    \textbf{NMF dim 29}  & -0.0664 & 0.049  & -1.352 & 0.176  & -0.163 & 0.030  \\
    \textbf{D.C. circuit}  & -0.7542 & 0.150  & -5.017 & 0.000  & -1.049 & -0.460 \\
    \textbf{circuit\_1}  & -0.4190 & 0.146  & -2.864 & 0.004  & -0.706 & -0.132 \\
    \textbf{circuit\_2}  & -0.7393 & 0.124  & -5.980 & 0.000  & -0.982 & -0.497 \\
    \textbf{circuit\_3}  & -0.6783 & 0.137  & -4.968 & 0.000  & -0.946 & -0.411 \\
    \textbf{circuit\_4}  & -0.7192 & 0.141  & -5.109 & 0.000  & -0.995 & -0.443 \\
    \textbf{circuit\_5}  & -0.4461 & 0.142  & -3.139 & 0.002  & -0.725 & -0.168 \\
    \textbf{circuit\_6}  & -0.5463 & 0.136  & -4.027 & 0.000  & -0.812 & -0.280 \\
    \textbf{circuit\_7}  & -0.6152 & 0.135  & -4.562 & 0.000  & -0.879 & -0.351 \\
    \textbf{circuit\_8}  & -0.1715 & 0.146  & -1.178 & 0.239  & -0.457 & 0.114  \\
    \textbf{circuit\_10} & -0.5374 & 0.152  & -3.537 & 0.000  & -0.835 & -0.240 \\
    \textbf{circuit\_11} & -0.4303 & 0.137  & -3.146 & 0.002  & -0.698 & -0.162 \\
    \textbf{decision\_date} & -0.2111 & 0.043  & -4.901 & 0.000  & -0.296 & -0.127 \\
    \midrule
    \textbf{Pseudo-R\textsuperscript{2}} & \textbf{6.2\%} & & & & & \\
    \bottomrule
    \end{tabular}

    \caption{\textbf{Logistic regression for civil rights cases.}}
    \label{tab:civil_rights_logistic_cite}
  \end{minipage}
  \hfill
  \begin{minipage}{0.49\textwidth}
    \centering
    \scriptsize
    
    \begin{tabular}{lcccccc}
    \toprule
    \textbf{judgement=1} & \textbf{coef} & \textbf{std err} & \textbf{z} & \textbf{P$>|$z$|$} & \textbf{[0.025} & \textbf{0.975]} \\
    \midrule
    \textbf{Intercept}    & -0.0393  & 0.083  & -0.475  & 0.635  & -0.202  & 0.123  \\
    \textbf{NMF dim 0}    & 0.0233   & 0.046  & 0.503   & 0.615  & -0.068  & 0.114  \\
    \textbf{NMF dim 1}    & -0.0122  & 0.030  & -0.408  & 0.683  & -0.071  & 0.046  \\
    \textbf{NMF dim 2}    & 0.0118   & 0.027  & 0.436   & 0.663  & -0.041  & 0.065  \\
    \textbf{NMF dim 3}    & -0.0906  & 0.049  & -1.855  & 0.064  & -0.186  & 0.005  \\
    \textbf{NMF dim 4}    & 0.0311   & 0.030  & 1.025   & 0.306  & -0.028  & 0.091  \\
    \textbf{NMF dim 5}    & -0.0310  & 0.024  & -1.274  & 0.203  & -0.079  & 0.017  \\
    \textbf{NMF dim 6}    & -0.0482  & 0.040  & -1.198  & 0.231  & -0.127  & 0.031  \\
    \textbf{NMF dim 7}    & -0.0489  & 0.029  & -1.714  & 0.087  & -0.105  & 0.007  \\
    \textbf{NMF dim 8}    & 0.0078   & 0.028  & 0.279   & 0.780  & -0.047  & 0.063  \\
    \textbf{NMF dim 9}    & -0.0248  & 0.021  & -1.183  & 0.237  & -0.066  & 0.016  \\
    \textbf{NMF dim 10}   & -0.0018  & 0.018  & -0.101  & 0.920  & -0.038  & 0.034  \\
    \textbf{NMF dim 11}   & 0.0030   & 0.023  & 0.132   & 0.895  & -0.041  & 0.047  \\
    \textbf{NMF dim 12}   & 0.1276   & 1.016  & 0.126   & 0.900  & -1.863  & 2.119  \\
    \textbf{NMF dim 13}   & 0.1281   & 0.987  & 0.130   & 0.897  & -1.807  & 2.063  \\
    \textbf{NMF dim 14}   & 0.0146   & 0.022  & 0.660   & 0.509  & -0.029  & 0.058  \\
    \textbf{NMF dim 15}   & -0.0525  & 0.025  & -2.066  & 0.039  & -0.102  & -0.003 \\
    \textbf{NMF dim 16}   & -0.0130  & 0.022  & -0.585  & 0.559  & -0.057  & 0.031  \\
    \textbf{NMF dim 17}   & 0.0187   & 0.022  & 0.850   & 0.395  & -0.024  & 0.062  \\
    \textbf{NMF dim 18}   & 0.0341   & 0.025  & 1.364   & 0.173  & -0.015  & 0.083  \\
    \textbf{NMF dim 19}   & 0.0082   & 0.022  & 0.366   & 0.715  & -0.036  & 0.052  \\
    \textbf{NMF dim 20}   & -0.0229  & 0.026  & -0.878  & 0.380  & -0.074  & 0.028  \\
    \textbf{NMF dim 21}   & -0.0226  & 0.037  & -0.614  & 0.540  & -0.095  & 0.049  \\
    \textbf{NMF dim 22}   & -0.0013  & 0.022  & -0.059  & 0.953  & -0.044  & 0.041  \\
    \textbf{NMF dim 23}   & -0.0152  & 0.024  & -0.645  & 0.519  & -0.061  & 0.031  \\
    \textbf{NMF dim 24}   & -0.0684  & 0.027  & -2.529  & 0.011  & -0.121  & -0.015 \\
    \textbf{NMF dim 25}   & -0.0167  & 0.024  & -0.707  & 0.480  & -0.063  & 0.030  \\
    \textbf{NMF dim 26}   & 0.0105   & 0.021  & 0.502   & 0.616  & -0.030  & 0.051  \\
    \textbf{NMF dim 27}   & 0.0142   & 0.025  & 0.576   & 0.565  & -0.034  & 0.062  \\
    \textbf{NMF dim 28}   & 0.0255   & 0.021  & 1.191   & 0.233  & -0.016  & 0.067  \\
    \textbf{NMF dim 29}   & 0.0613   & 0.027  & 2.301   & 0.021  & 0.009   & 0.113  \\
    \textbf{D.C. circuit}   & -0.1079  & 0.145  & -0.742  & 0.458  & -0.393  & 0.177  \\
    \textbf{circuit\_1}   & -0.1042  & 0.116  & -0.898  & 0.369  & -0.332  & 0.123  \\
    \textbf{circuit\_2}   & -0.0462  & 0.094  & -0.492  & 0.623  & -0.230  & 0.138  \\
    \textbf{circuit\_3}   & 0.0797   & 0.101  & 0.787   & 0.431  & -0.119  & 0.278  \\
    \textbf{circuit\_4}   & 0.0753   & 0.110  & 0.681   & 0.496  & -0.141  & 0.292  \\
    \textbf{circuit\_5}   & 0.0618   & 0.108  & 0.571   & 0.568  & -0.150  & 0.274  \\
    \textbf{circuit\_6}   & 0.1224   & 0.113  & 1.082   & 0.279  & -0.099  & 0.344  \\
    \textbf{circuit\_7}   & -0.2011  & 0.121  & -1.666  & 0.096  & -0.438  & 0.036  \\
    \textbf{circuit\_8}   & 0.1660   & 0.112  & 1.488   & 0.137  & -0.053  & 0.385  \\
    \textbf{circuit\_10}  & 0.2049   & 0.125  & 1.644   & 0.100  & -0.039  & 0.449  \\
    \textbf{circuit\_11}  & 0.1993   & 0.114  & 1.749   & 0.080  & -0.024  & 0.423  \\
    \textbf{decision\_date}& -0.1344  & 0.046  & -2.929  & 0.003  & -0.224  & -0.044 \\
    \midrule
    \textbf{Pseudo-R\textsuperscript{2}} & \textbf{1.6\%} & & & & & \\
    \bottomrule
    \end{tabular}

    \caption{\textbf{Logistic regression for contract cases.}}
    \label{tab:contract_logistic_cite}
  \end{minipage}
\end{figure}

\begin{figure}[!ht]
  \centering
  \begin{minipage}{0.49\textwidth}
    \centering
    \scriptsize
    
    \begin{tabular}{lcccccc}
    \toprule
    \textbf{judgement=1} & \textbf{coef} & \textbf{std err} & \textbf{z} & \textbf{P$>|$z$|$} & \textbf{[0.025} & \textbf{0.975]} \\
    \midrule
    \textbf{Intercept}    & 0.6652  & 0.151  & 4.413  & 0.000  & 0.370  & 0.961  \\
    \textbf{NMF dim 0}    & -0.0397 & 0.105  & -0.380 & 0.704  & -0.245 & 0.165  \\
    \textbf{NMF dim 1}    & -0.5092 & 0.103  & -4.952 & 0.000  & -0.711 & -0.308 \\
    \textbf{NMF dim 2}    & -0.0597 & 0.054  & -1.101 & 0.271  & -0.166 & 0.047  \\
    \textbf{NMF dim 3}    & -0.0333 & 0.049  & -0.678 & 0.498  & -0.129 & 0.063  \\
    \textbf{NMF dim 4}    & -0.3172 & 0.104  & -3.051 & 0.002  & -0.521 & -0.113 \\
    \textbf{NMF dim 5}    & -0.1640 & 0.072  & -2.275 & 0.023  & -0.305 & -0.023 \\
    \textbf{NMF dim 6}    & -0.0352 & 0.057  & -0.615 & 0.538  & -0.147 & 0.077  \\
    \textbf{NMF dim 7}    & -0.1782 & 0.063  & -2.838 & 0.005  & -0.301 & -0.055 \\
    \textbf{NMF dim 8}    & -0.1147 & 0.069  & -1.666 & 0.096  & -0.250 & 0.020  \\
    \textbf{NMF dim 9}    & -0.3096 & 0.070  & -4.393 & 0.000  & -0.448 & -0.171 \\
    \textbf{NMF dim 10}   & -0.2047 & 0.058  & -3.555 & 0.000  & -0.318 & -0.092 \\
    \textbf{NMF dim 11}   & -0.0531 & 0.052  & -1.030 & 0.303  & -0.154 & 0.048  \\
    \textbf{NMF dim 12}   & -0.1283 & 0.068  & -1.884 & 0.060  & -0.262 & 0.005  \\
    \textbf{NMF dim 13}   & -0.0271 & 0.060  & -0.449 & 0.653  & -0.145 & 0.091  \\
    \textbf{NMF dim 14}   & -0.1045 & 0.065  & -1.611 & 0.107  & -0.232 & 0.023  \\
    \textbf{NMF dim 15}   & -0.4046 & 0.082  & -4.907 & 0.000  & -0.566 & -0.243 \\
    \textbf{NMF dim 16}   & -0.0573 & 0.076  & -0.757 & 0.449  & -0.205 & 0.091  \\
    \textbf{NMF dim 17}   & -0.1048 & 0.060  & -1.756 & 0.079  & -0.222 & 0.012  \\
    \textbf{NMF dim 18}   & -0.1514 & 0.064  & -2.368 & 0.018  & -0.277 & -0.026 \\
    \textbf{NMF dim 19}   & 0.1830  & 0.096  & 1.898  & 0.058  & -0.006 & 0.372  \\
    \textbf{NMF dim 20}   & -0.1856 & 0.077  & -2.412 & 0.016  & -0.336 & -0.035 \\
    \textbf{NMF dim 21}   & -0.0885 & 0.059  & -1.491 & 0.136  & -0.205 & 0.028  \\
    \textbf{NMF dim 22}   & 0.0810  & 0.208  & 0.390  & 0.697  & -0.326 & 0.488  \\
    \textbf{NMF dim 23}   & -0.0840 & 0.057  & -1.468 & 0.142  & -0.196 & 0.028  \\
    \textbf{NMF dim 24}   & -0.0542 & 0.071  & -0.764 & 0.445  & -0.193 & 0.085  \\
    \textbf{NMF dim 25}   & -0.2426 & 0.066  & -3.697 & 0.000  & -0.371 & -0.114 \\
    \textbf{NMF dim 26}   & 0.0659  & 0.075  & 0.874  & 0.382  & -0.082 & 0.214  \\
    \textbf{NMF dim 27}   & -0.1422 & 0.064  & -2.224 & 0.026  & -0.268 & -0.017 \\
    \textbf{NMF dim 28}   & -0.1770 & 0.084  & -2.112 & 0.035  & -0.341 & -0.013 \\
    \textbf{NMF dim 29}   & -0.0952 & 0.056  & -1.712 & 0.087  & -0.204 & 0.014  \\
    \textbf{D.C. circuit}   & -1.6609 & 0.303  & -5.483 & 0.000  & -2.255 & -1.067 \\
    \textbf{circuit\_1}   & -1.0070 & 0.230  & -4.386 & 0.000  & -1.457 & -0.557 \\
    \textbf{circuit\_2}   & -1.1082 & 0.188  & -5.890 & 0.000  & -1.477 & -0.739 \\
    \textbf{circuit\_3}   & -1.2574 & 0.209  & -6.009 & 0.000  & -1.667 & -0.847 \\
    \textbf{circuit\_4}   & -0.9900 & 0.226  & -4.372 & 0.000  & -1.434 & -0.546 \\
    \textbf{circuit\_5}   & -0.2812 & 0.239  & -1.175 & 0.240  & -0.750 & 0.188  \\
    \textbf{circuit\_6}   & -0.5400 & 0.217  & -2.483 & 0.013  & -0.966 & -0.114 \\
    \textbf{circuit\_7}   & -0.6394 & 0.231  & -2.774 & 0.006  & -1.091 & -0.188 \\
    \textbf{circuit\_8}   & -0.5307 & 0.245  & -2.169 & 0.030  & -1.010 & -0.051 \\
    \textbf{circuit\_10}  & -1.0166 & 0.256  & -3.976 & 0.000  & -1.518 & -0.515 \\
    \textbf{circuit\_11}  & -0.4355 & 0.268  & -1.628 & 0.104  & -0.960 & 0.089  \\
    \textbf{decision\_date}& 0.1223  & 0.093  & 1.317  & 0.188  & -0.060 & 0.304  \\
    \midrule
    \textbf{Pseudo-R\textsuperscript{2}} & \textbf{8.9\%} & & & & & \\
    \bottomrule
\end{tabular}

    \caption{\textbf{Logistic regression for prisoner petition cases.}}
    \label{tab:prisoner_petitions_logistic_cite}
  \end{minipage}
  \hfill
  \begin{minipage}{0.49\textwidth}
    \centering
    \scriptsize
    
    \begin{tabular}{lcccccc}
    \toprule
    \textbf{judgement=1} & \textbf{coef} & \textbf{std err} & \textbf{z} & \textbf{P$>|$z$|$} & \textbf{[0.025} & \textbf{0.975]} \\
    \midrule
    \textbf{Intercept}    & -0.2482  & 0.142  & -1.750  & 0.080  & -0.526  & 0.030  \\
    \textbf{NMF dim 0}    & -0.1561  & 0.078  & -2.001  & 0.045  & -0.309  & -0.003 \\
    \textbf{NMF dim 1}    & 0.0123   & 0.027  & 0.463   & 0.644  & -0.040  & 0.065  \\
    \textbf{NMF dim 2}    & -0.0983  & 0.050  & -1.981  & 0.048  & -0.196  & -0.001 \\
    \textbf{NMF dim 3}    & -0.2335  & 0.082  & -2.864  & 0.004  & -0.393  & -0.074 \\
    \textbf{NMF dim 4}    & -0.0315  & 0.037  & -0.853  & 0.393  & -0.104  & 0.041  \\
    \textbf{NMF dim 5}    & 0.0034   & 0.030  & 0.113   & 0.910  & -0.055  & 0.062  \\
    \textbf{NMF dim 6}    & 0.0427   & 0.031  & 1.357   & 0.175  & -0.019  & 0.104  \\
    \textbf{NMF dim 7}    & -0.0940  & 0.037  & -2.539  & 0.011  & -0.167  & -0.021 \\
    \textbf{NMF dim 8}    & 0.0097   & 0.046  & 0.213   & 0.831  & -0.080  & 0.099  \\
    \textbf{NMF dim 9}    & 0.0373   & 0.035  & 1.066   & 0.287  & -0.031  & 0.106  \\
    \textbf{NMF dim 10}   & -0.1489  & 0.054  & -2.759  & 0.006  & -0.255  & -0.043 \\
    \textbf{NMF dim 11}   & -0.0498  & 0.029  & -1.712  & 0.087  & -0.107  & 0.007  \\
    \textbf{NMF dim 12}   & -0.0058  & 0.032  & -0.181  & 0.857  & -0.069  & 0.057  \\
    \textbf{NMF dim 13}   & 0.1601   & 3.524  & 0.045   & 0.964  & -6.746  & 7.066  \\
    \textbf{NMF dim 14}   & 0.1066   & 0.035  & 3.074   & 0.002  & 0.039   & 0.175  \\
    \textbf{NMF dim 15}   & 0.0090   & 0.035  & 0.257   & 0.797  & -0.059  & 0.077  \\
    \textbf{NMF dim 16}   & -0.0040  & 0.027  & -0.149  & 0.882  & -0.056  & 0.048  \\
    \textbf{NMF dim 17}   & -0.0054  & 0.029  & -0.189  & 0.850  & -0.062  & 0.051  \\
    \textbf{NMF dim 18}   & -0.0023  & 0.047  & -0.049  & 0.961  & -0.094  & 0.089  \\
    \textbf{NMF dim 19}   & -0.0060  & 0.031  & -0.194  & 0.846  & -0.067  & 0.055  \\
    \textbf{NMF dim 20}   & 0.0124   & 0.033  & 0.372   & 0.710  & -0.053  & 0.078  \\
    \textbf{NMF dim 21}   & -0.1117  & 0.044  & -2.549  & 0.011  & -0.198  & -0.026 \\
    \textbf{NMF dim 22}   & 0.0810   & 0.035  & 2.299   & 0.021  & 0.012   & 0.150  \\
    \textbf{NMF dim 23}   & -0.0184  & 0.026  & -0.699  & 0.484  & -0.070  & 0.033  \\
    \textbf{NMF dim 24}   & -0.0721  & 0.041  & -1.762  & 0.078  & -0.152  & 0.008  \\
    \textbf{NMF dim 25}   & 0.0011   & 0.023  & 0.049   & 0.961  & -0.044  & 0.047  \\
    \textbf{NMF dim 26}   & 0.0399   & 0.035  & 1.143   & 0.253  & -0.028  & 0.108  \\
    \textbf{NMF dim 27}   & -0.0149  & 0.042  & -0.359  & 0.720  & -0.096  & 0.066  \\
    \textbf{NMF dim 28}   & -0.0469  & 0.058  & -0.805  & 0.421  & -0.161  & 0.067  \\
    \textbf{NMF dim 29}   & -0.2647  & 0.300  & -0.882  & 0.378  & -0.853  & 0.324  \\
    \textbf{D.C. circuit}   & 0.9718   & 0.191  & 5.081   & 0.000  & 0.597   & 1.347  \\
    \textbf{circuit\_1}   & 0.2232   & 0.181  & 1.231   & 0.218  & -0.132  & 0.578  \\
    \textbf{circuit\_2}   & -0.0438  & 0.150  & -0.292  & 0.771  & -0.338  & 0.251  \\
    \textbf{circuit\_3}   & 0.2454   & 0.155  & 1.583   & 0.113  & -0.058  & 0.549  \\
    \textbf{circuit\_4}   & -0.0203  & 0.174  & -0.117  & 0.907  & -0.361  & 0.320  \\
    \textbf{circuit\_5}   & 0.4053   & 0.155  & 2.620   & 0.009  & 0.102   & 0.708  \\
    \textbf{circuit\_6}   & -0.1176  & 0.172  & -0.684  & 0.494  & -0.455  & 0.219  \\
    \textbf{circuit\_7}   & -0.1836  & 0.204  & -0.898  & 0.369  & -0.584  & 0.217  \\
    \textbf{circuit\_8}   & 0.3364   & 0.180  & 1.870   & 0.062  & -0.016  & 0.689  \\
    \textbf{circuit\_10}  & 0.1063   & 0.213  & 0.499   & 0.618  & -0.311  & 0.524  \\
    \textbf{circuit\_11}  & 0.0040   & 0.184  & 0.021   & 0.983  & -0.357  & 0.365  \\
    \textbf{decision\_date}& -0.0852  & 0.063  & -1.353  & 0.176  & -0.209  & 0.038  \\
    \midrule
    \textbf{Pseudo-R\textsuperscript{2}} & \textbf{6.6\%} & & & & & \\
    \bottomrule
\end{tabular}

    \caption{\textbf{Logistic regression for tort cases.}}
    \label{tab:tort_logistic_cite}
  \end{minipage}
\end{figure}

\begin{figure}[!ht]
  \centering
  \begin{minipage}{0.49\textwidth}
    \centering
    \scriptsize

    \begin{tabular}{lcccccc}
    \toprule
    \textbf{judgement=1} & \textbf{coef} & \textbf{std err} & \textbf{z} & \textbf{P$>|$z$|$} & \textbf{[0.025} & \textbf{0.975]} \\
    \midrule
    \textbf{Intercept}    & 0.1956   & 0.106  & 1.850   & 0.064  & -0.012  & 0.403  \\
    \textbf{NMF dim 0}    & -0.1101  & 0.066  & -1.676  & 0.094  & -0.239  & 0.019  \\
    \textbf{NMF dim 1}    & -0.0363  & 0.040  & -0.909  & 0.363  & -0.115  & 0.042  \\
    \textbf{NMF dim 2}    & -0.0906  & 0.070  & -1.288  & 0.198  & -0.228  & 0.047  \\
    \textbf{NMF dim 3}    & -0.0274  & 0.037  & -0.747  & 0.455  & -0.099  & 0.044  \\
    \textbf{NMF dim 4}    & 0.0605   & 0.042  & 1.427   & 0.153  & -0.023  & 0.144  \\
    \textbf{NMF dim 5}    & -0.0602  & 0.033  & -1.837  & 0.066  & -0.124  & 0.004  \\
    \textbf{NMF dim 6}    & 0.0292   & 0.042  & 0.688   & 0.491  & -0.054  & 0.113  \\
    \textbf{NMF dim 7}    & -0.0072  & 0.034  & -0.209  & 0.834  & -0.075  & 0.060  \\
    \textbf{NMF dim 8}    & -0.0216  & 0.034  & -0.638  & 0.524  & -0.088  & 0.045  \\
    \textbf{NMF dim 9}    & 0.0288   & 0.033  & 0.864   & 0.388  & -0.037  & 0.094  \\
    \textbf{NMF dim 10}   & 0.0403   & 0.027  & 1.490   & 0.136  & -0.013  & 0.093  \\
    \textbf{NMF dim 11}   & 0.0125   & 0.042  & 0.297   & 0.766  & -0.070  & 0.095  \\
    \textbf{NMF dim 12}   & 0.0376   & 0.040  & 0.940   & 0.347  & -0.041  & 0.116  \\
    \textbf{NMF dim 13}   & -0.0273  & 0.034  & -0.800  & 0.424  & -0.094  & 0.040  \\
    \textbf{NMF dim 14}   & -0.0481  & 0.036  & -1.349  & 0.177  & -0.118  & 0.022  \\
    \textbf{NMF dim 15}   & -0.0568  & 0.033  & -1.734  & 0.083  & -0.121  & 0.007  \\
    \textbf{NMF dim 16}   & -0.0190  & 0.047  & -0.406  & 0.685  & -0.110  & 0.073  \\
    \textbf{NMF dim 17}   & -0.0740  & 0.042  & -1.777  & 0.076  & -0.156  & 0.008  \\
    \textbf{NMF dim 18}   & 0.0353   & 0.046  & 0.770   & 0.441  & -0.055  & 0.125  \\
    \textbf{NMF dim 19}   & -0.0014  & 0.028  & -0.049  & 0.961  & -0.057  & 0.054  \\
    \textbf{NMF dim 20}   & -0.0040  & 0.042  & -0.096  & 0.923  & -0.086  & 0.078  \\
    \textbf{NMF dim 21}   & 0.0063   & 0.026  & 0.238   & 0.812  & -0.045  & 0.058  \\
    \textbf{NMF dim 22}   & -0.0502  & 0.037  & -1.357  & 0.175  & -0.123  & 0.022  \\
    \textbf{NMF dim 23}   & 0.0851   & 0.035  & 2.431   & 0.015  & 0.016   & 0.154  \\
    \textbf{NMF dim 24}   & 0.0363   & 0.033  & 1.106   & 0.269  & -0.028  & 0.101  \\
    \textbf{NMF dim 25}   & -0.0094  & 0.048  & -0.198  & 0.843  & -0.103  & 0.084  \\
    \textbf{NMF dim 26}   & 0.0224   & 0.031  & 0.719   & 0.472  & -0.039  & 0.084  \\
    \textbf{NMF dim 27}   & 0.0322   & 0.032  & 1.001   & 0.317  & -0.031  & 0.095  \\
    \textbf{NMF dim 28}   & -0.0203  & 0.036  & -0.559  & 0.576  & -0.092  & 0.051  \\
    \textbf{NMF dim 29}   & -0.0235  & 0.035  & -0.681  & 0.496  & -0.091  & 0.044  \\
    \textbf{D.C. circuit}   & 0.2087   & 0.167  & 1.249   & 0.212  & -0.119  & 0.536  \\
    \textbf{circuit\_1}   & -0.5952  & 0.159  & -3.736  & 0.000  & -0.908  & -0.283  \\
    \textbf{circuit\_2}   & -0.2019  & 0.129  & -1.569  & 0.117  & -0.454  & 0.050  \\
    \textbf{circuit\_3}   & -0.1317  & 0.141  & -0.935  & 0.350  & -0.408  & 0.144  \\
    \textbf{circuit\_4}   & -0.2154  & 0.151  & -1.428  & 0.153  & -0.511  & 0.080  \\
    \textbf{circuit\_5}   & -0.1880  & 0.159  & -1.186  & 0.236  & -0.499  & 0.123  \\
    \textbf{circuit\_6}   & -0.4113  & 0.137  & -2.996  & 0.003  & -0.680  & -0.142  \\
    \textbf{circuit\_7}   & -0.0966  & 0.148  & -0.654  & 0.513  & -0.386  & 0.193  \\
    \textbf{circuit\_8}   & -0.2360  & 0.149  & -1.588  & 0.112  & -0.527  & 0.055  \\
    \textbf{circuit\_10}  & -0.3923  & 0.190  & -2.067  & 0.039  & -0.764  & -0.020  \\
    \textbf{circuit\_11}  & -0.1751  & 0.152  & -1.154  & 0.248  & -0.472  & 0.122  \\
    \textbf{decision\_date}& 0.1241   & 0.062  & 1.998   & 0.046  & 0.002   & 0.246  \\
    \midrule
    \textbf{Pseudo-R\textsuperscript{2}} & \textbf{1.4\%} & & & & & \\
    \bottomrule
    \end{tabular}

    \caption{\textbf{Logistic regression for labor cases.}}
    \label{tab:labor_logistic_cite}
  \end{minipage}
  \hfill
  \begin{minipage}{0.49\textwidth}
    \centering
    \scriptsize

    \begin{tabular}{lcccccc}
    \toprule
    \textbf{judgement=1} & \textbf{coef} & \textbf{std err} & \textbf{z} & \textbf{P$>|$z$|$} & \textbf{[0.025} & \textbf{0.975]} \\
    \midrule
    \textbf{Intercept}    & 0.0485   & 0.051  & 0.959   & 0.337  & -0.051  & 0.147  \\
    \textbf{NMF dim 0}    & 0.0021   & 0.033  & 0.065   & 0.948  & -0.062  & 0.067  \\
    \textbf{NMF dim 1}    & -0.0636  & 0.024  & -2.689  & 0.007  & -0.110  & -0.017 \\
    \textbf{NMF dim 2}    & 0.0082   & 0.018  & 0.451   & 0.652  & -0.027  & 0.044  \\
    \textbf{NMF dim 3}    & -0.0314  & 0.024  & -1.299  & 0.194  & -0.079  & 0.016  \\
    \textbf{NMF dim 4}    & -0.0872  & 0.035  & -2.494  & 0.013  & -0.156  & -0.019 \\
    \textbf{NMF dim 5}    & 0.0267   & 0.016  & 1.717   & 0.086  & -0.004  & 0.057  \\
    \textbf{NMF dim 6}    & -0.0171  & 0.017  & -1.014  & 0.311  & -0.050  & 0.016  \\
    \textbf{NMF dim 7}    & -0.0103  & 0.017  & -0.591  & 0.555  & -0.044  & 0.024  \\
    \textbf{NMF dim 8}    & 0.0255   & 0.022  & 1.169   & 0.243  & -0.017  & 0.068  \\
    \textbf{NMF dim 9}    & 0.0054   & 0.017  & 0.329   & 0.742  & -0.027  & 0.038  \\
    \textbf{NMF dim 10}   & -0.1165  & 0.021  & -5.494  & 0.000  & -0.158  & -0.075 \\
    \textbf{NMF dim 11}   & -0.0112  & 0.018  & -0.624  & 0.533  & -0.046  & 0.024  \\
    \textbf{NMF dim 12}   & 0.1064   & 0.049  & 2.174   & 0.030  & 0.010   & 0.202  \\
    \textbf{NMF dim 13}   & 0.0239   & 0.018  & 1.356   & 0.175  & -0.011  & 0.058  \\
    \textbf{NMF dim 14}   & 0.0095   & 0.016  & 0.607   & 0.544  & -0.021  & 0.040  \\
    \textbf{NMF dim 15}   & 0.0328   & 0.017  & 1.892   & 0.058  & -0.001  & 0.067  \\
    \textbf{NMF dim 16}   & 0.0493   & 0.017  & 2.818   & 0.005  & 0.015   & 0.084  \\
    \textbf{NMF dim 17}   & -0.0062  & 0.021  & -0.300  & 0.764  & -0.047  & 0.035  \\
    \textbf{NMF dim 18}   & 0.0149   & 0.017  & 0.867   & 0.386  & -0.019  & 0.049  \\
    \textbf{NMF dim 19}   & 0.1083   & 0.023  & 4.802   & 0.000  & 0.064   & 0.152  \\
    \textbf{NMF dim 20}   & 0.0389   & 0.025  & 1.526   & 0.127  & -0.011  & 0.089  \\
    \textbf{NMF dim 21}   & -0.0139  & 0.016  & -0.876  & 0.381  & -0.045  & 0.017  \\
    \textbf{NMF dim 22}   & 0.0131   & 0.017  & 0.790   & 0.430  & -0.019  & 0.046  \\
    \textbf{NMF dim 23}   & 0.0192   & 0.017  & 1.115   & 0.265  & -0.015  & 0.053  \\
    \textbf{NMF dim 24}   & 0.0207   & 0.019  & 1.079   & 0.281  & -0.017  & 0.058  \\
    \textbf{NMF dim 25}   & 0.0365   & 0.020  & 1.865   & 0.062  & -0.002  & 0.075  \\
    \textbf{NMF dim 26}   & 0.0381   & 0.017  & 2.284   & 0.022  & 0.005   & 0.071  \\
    \textbf{NMF dim 27}   & -0.0112  & 0.018  & -0.617  & 0.537  & -0.047  & 0.024  \\
    \textbf{NMF dim 28}   & 0.0087   & 0.016  & 0.535   & 0.592  & -0.023  & 0.041  \\
    \textbf{NMF dim 29}   & 0.0037   & 0.016  & 0.225   & 0.822  & -0.028  & 0.036  \\
    \textbf{D.C. circuit}   & -0.5941  & 0.080  & -7.448  & 0.000  & -0.751  & -0.438 \\
    \textbf{circuit\_1}   & -0.0401  & 0.080  & -0.498  & 0.618  & -0.198  & 0.118  \\
    \textbf{circuit\_2}   & -0.0965  & 0.063  & -1.544  & 0.123  & -0.219  & 0.026  \\
    \textbf{circuit\_3}   & -0.0461  & 0.069  & -0.666  & 0.506  & -0.182  & 0.090  \\
    \textbf{circuit\_4}   & 0.1050   & 0.079  & 1.322   & 0.186  & -0.051  & 0.261  \\
    \textbf{circuit\_5}   & -0.1597  & 0.085  & -1.870  & 0.061  & -0.327  & 0.008  \\
    \textbf{circuit\_6}   & -0.1109  & 0.078  & -1.421  & 0.155  & -0.264  & 0.042  \\
    \textbf{circuit\_7}   & -0.0715  & 0.080  & -0.895  & 0.371  & -0.228  & 0.085  \\
    \textbf{circuit\_8}   & 0.2236   & 0.082  & 2.722   & 0.006  & 0.063   & 0.385  \\
    \textbf{circuit\_10}  & 0.1238   & 0.091  & 1.361   & 0.174  & -0.055  & 0.302  \\
    \textbf{circuit\_11}  & 0.2416   & 0.082  & 2.960   & 0.003  & 0.082   & 0.402  \\
    \textbf{decision\_date}& -0.0174  & 0.034  & -0.507  & 0.612  & -0.085  & 0.050  \\
    \midrule
    \textbf{Pseudo-R\textsuperscript{2}} & \textbf{1.9\%} & & & & & \\
    \bottomrule
    \end{tabular}

    \caption{\textbf{Logistic regression for ``other'' cases.}}
    \label{tab:other_logistic_cite}
  \end{minipage}
\end{figure}

\clearpage
\newpage
\section{Judge Predictability: Sensitivity Analysis}
\label{SI:judge_predictability_sensitivity}

In the final section of our main paper, we analyze the predictability of case outcomes on a per-judge basis. 
Specifically, we test the null hypothesis for each judge asserting that the case outcome is not predictable beyond what would be expected from random guessing. 
To ensure the representativeness of these hypothesis tests, we only include judges for whom we have at least 50 cases after down-sampling the majority class (i.e., for each judge, we require a minimum of 25 test cases in which the plaintiff won and 25 test cases in which the plaintiff lost).

We also need to determine the confidence level at which we consider the gradient boosting model’s outcome prediction to be valid. 
This is conceptually similar to the analysis in \ref{SI:transformer_labeling}, where we considered the confidence of the transformer model on the comparatively simpler task of predicting the outcome based on the judge’s case summary. 
Here, we perform a similar analysis based on the predictions of the gradient boosting model, denoted by $q \in [0,1]$. 
In direct analogy to the classification confidence $\tau$ relative to the transformer prediction $p$, 
we define the gradient boost model’s prediction confidence level $\kappa$ as $|q - 0.5|$.

\begin{figure*}[!htb]
    \centering
    \includegraphics[width=0.8\linewidth]{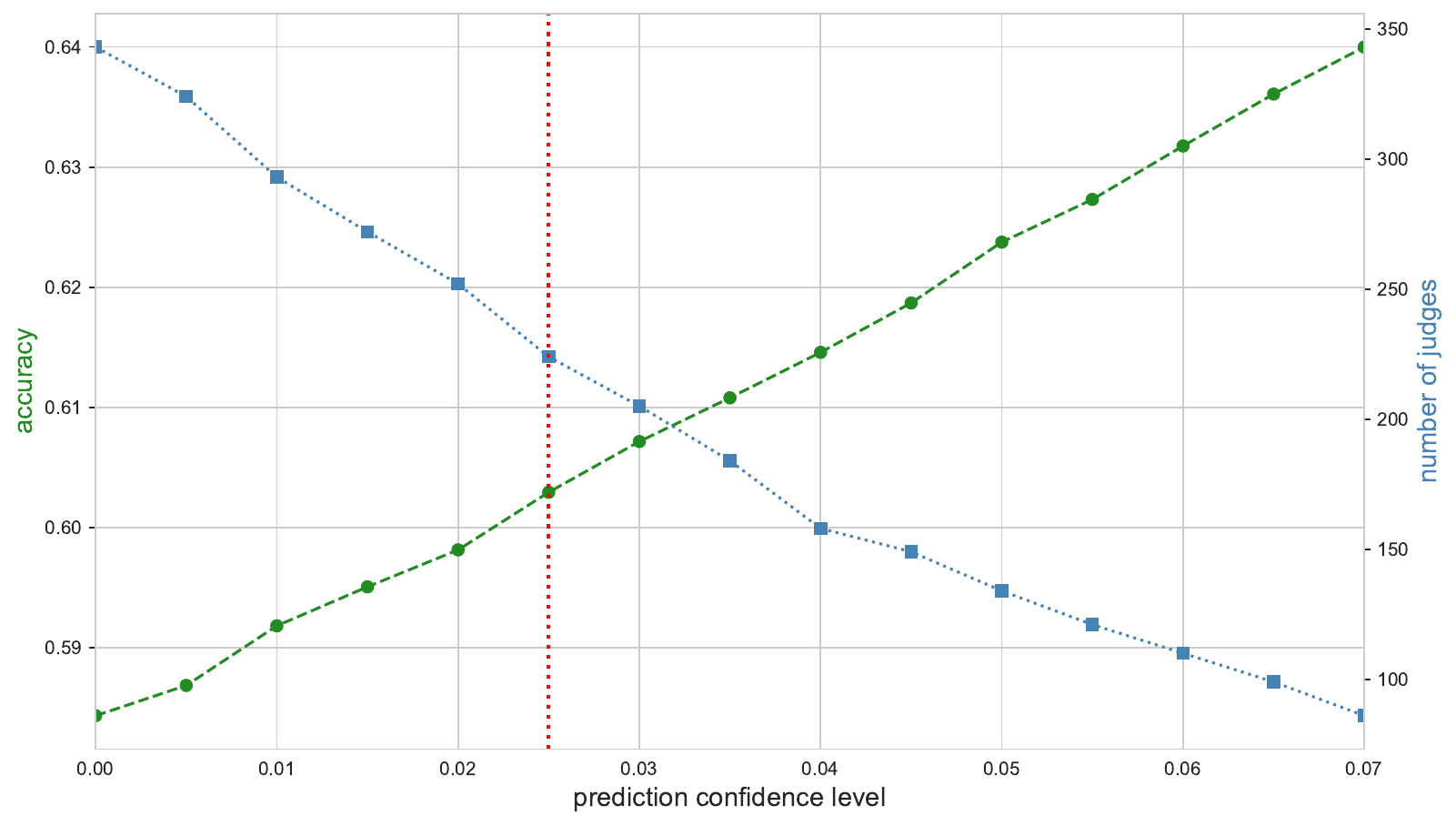}
    \caption{\textbf{Trade-off between GB model threshold and number of judges in the data.} As we increase the prediction cutoff $\kappa$, the number of judges in our data shrinks.}
    \label{fig:case_predictability_confidence}
\end{figure*}

In Figure \ref{fig:case_predictability_confidence}, we show the overall prediction accuracy as a function of $\kappa$. 
(Note that the range of $\kappa$ is much narrower than that of $\tau$, which is expected since this classification task is more challenging, leading to lower confidence in the model’s predictions.) 
As anticipated, and consistent with our observations in Figure 2 of the main paper, we find that the larger $\kappa$, the higher the prediction accuracy on the test set. 
However, as $\kappa$ increases, the number of cases decreases, resulting in fewer judges with more than 50 cases.
For the presentation of our results in Figure 5 of the main paper, we thus set $\kappa = 0.025$ as a good trade-off, which leaves us with 224 judges.

Equipped with case outcome predictions across 224 judges, we can then test the null hypothesis of random guessing for each judge. 
Following our discussion in \ref{SI:p-value_primer}, we adjust the 224 p-values using the Benjamini-Hochberg (BH) correction, as the samples are independent of one another. 
For the example shown in Figure 5 of the main paper, the null hypothesis is rejected for $7.4\%$ of judges, with 22 out of 224 judges found to be predictable at a confidence level exceeding $90\%$.

However, there is an additional source of stochasticity that affects these results: 
the balanced down-sampling of cases. 
Consider a situation where, for a given judge, we observe 32 cases in which the plaintiff won and 25 cases in which the plaintiff lost. In this scenario, there are $(32,7)=3,365,856$ possible ways to down-sample to a balanced test set with 25 wins and 25 losses. 
To account for this stochasticity, we repeat the analysis from Figure 5 thirty times and compute the mean and standard deviations.
On average, $6.9\%$ of judges are found to be predictable beyond what would be expected from random guessing, with a standard deviation of $2.3\%$ after applying the BH correction. 
A more detailed overview is provided in Table \ref{tab:judge_predictability_stats}. We also repeat the entire analysis using the first 20\% of judges' decision history to calculate citation representations and find qualitatively similar results, see Table~\ref{tab:judge_predictability_stats_20}

\begin{table}[ht]
\centering
\begin{tabular}{|l|c|c|c|c|c|c|c|}
\hline
\textbf{Metric} & \textbf{Mean} & \textbf{Std} & \textbf{Min} & \textbf{25\%} & \textbf{50\%} & \textbf{75\%} & \textbf{Max} \\
\hline
judge predictability undecided & 108 & 2 & 105 & 107 & 108 & 109 & 113 \\
judge predictability outperforming & 23 & 2 & 18 & 22 & 23 & 24 & 26 \\
judge predictability underperforming & 2 & 1 & 1 & 2 & 2 & 2 & 4 \\
fraction of p-values below 5\% (raw) & 17.1\% & 1.3\% & 13.5\% & 16.5\% & 17.3\% & 18.1\% & 19.5\% \\
fraction of p-values below 5\% (BF adjusted) & 3.4\% & 0.8\% & 2.3\% & 3.0\% & 3.8\% & 3.8\% & 6.0\% \\
fraction of p-values below 5\% (BH adjusted) & 6.9\% & 0.9\% & 5.3\% & 6.0\% & 6.8\% & 7.5\% & 9.0\% \\
fraction of p-values below 5\% (BY adjusted) & 3.3\% & 1.0\% & 1.5\% & 2.3\% & 3.8\% & 3.8\% & 6.0\% \\
\hline
\end{tabular}
\caption{\textbf{Comparison of different p-value corrections.} We use the first 10\% cases assigned to a judge to construct the embeddings (as in the main paper). No matter what correction is selected, we find a small but significant set of judges to be predictable.}
\label{tab:judge_predictability_stats}
\end{table}

\begin{table}[ht]
\centering
\begin{tabular}{|l|c|c|c|c|c|c|c|}
\hline
\textbf{Metric} & \textbf{Mean} & \textbf{Std} & \textbf{Min} & \textbf{25\%} & \textbf{50\%} & \textbf{75\%} & \textbf{Max} \\
\hline
judge predictability undecided & 103 & 2 & 99 & 103 & 104 & 104.75 & 107 \\
judge predictability outperforming & 20 & 1 & 17 & 19 & 20 & 21 & 22 \\
judge predictability underperforming & 2 & 1 & 1 & 1 & 2 & 2 & 4 \\
fraction of p-values below 5\% (raw) & 15.97\% & 1.10\% & 13.6\% & 15.2\% & 16.0\% & 16.8\% & 17.6\% \\
fraction of p-values below 5\% (BF adjusted) & 4.1\% & 0.2\% & 4.0\% & 4.0\% & 4.0\% & 4.0\% & 4.8\% \\
fraction of p-values below 5\% (BH adjusted) & 5.8\% & 0.8\% & 4.0\% & 5.0\% & 5.6\% & 6.4\% & 7.2\% \\
fraction of p-values below 5\% (BY adjusted) & 4.1\% & 0.2\% & 4.0\% & 4.0\% & 4.0\% & 4.0\% & 4.8\% \\
\hline
\end{tabular}
\caption{\textbf{Comparison of different p-value corrections.} We use the first 20\% cases assigned to a judge to construct the embeddings (as in the main paper). As in the case of Table~\ref{tab:judge_predictability_stats}, no matter what correction is selected, we find a small but significant set of judges to be predictable.}
\label{tab:judge_predictability_stats_20}
\end{table}

\end{document}